\documentclass{WileyMSP-template}

\usepackage{amsmath}
\usepackage{graphicx}
\usepackage{microtype}
\usepackage{subcaption}
\usepackage{xcolor}
\usepackage{braket}
\usepackage{epsfig}
\usepackage{xprintlen}
\usepackage{pgffor}
\usepackage{booktabs}
\usepackage{tabularx}
\usepackage{soul}
\usepackage[comma,super,sort&compress,square]{natbib}

\newcommand{\be}{\begin{equation}}
\newcommand{\ee}{\end{equation}}
\newcommand{\bea}{\begin{eqnarray}}
\newcommand{\eea}{\end{eqnarray}}

\newcommand{\ws}{{WS$_2$}}
\newcommand{\mos}{{MoS$_2$}}
\newcommand{\mow}{Mo$_\mathrm{W}$}
\newcommand{\ch}{$\mathrm{(CH_2)_S}$}
\newcommand{\dd}[1]{D$_{#1}$}

\def\qq		{{\bf q}}

\def\fbf {5$\times$5~}

\def\by {$\times$}

\def\bverb      {\renewcommand{\baselinestretch}{1}\begin{verbatim}}
\def\everb  	{\end{verbatim}\renewcommand{\baselinestretch}{1.3}}

\def\qq         {{\mathbf q}}

\newcommand{\etsf} {European Theoretical Spectroscopy Facility (ETSF) www.etsf.eu}
\newcommand{\liege}{nanomat/Q-mat/CESAM, Universit\'e de Li\`ege, Institut de Physique, B-4000 Sart Tilman, Li\`ege, Belgium}
\newcommand{\barcelona}{Catalan Institute of Nanoscience and Nanotechnology (ICN2), CSIC and BIST, Campus UAB, Bellaterra, 08193 Barcelona, Spain}
\newcommand{\utrecht}{Chemistry Department, Debye Institute for Nanomaterials Science, Condensed Matter and Interfaces, Utrecht University, PO Box 80.000, 3508 TA Utrecht, The Netherlands}

\newcommand{\yambo} {{\normalfont\ttfamily Yambo}~}

\begin{document}

\pagestyle{fancy}

\title{Optical Signatures of Defect Centres in Transition Metal Dichalcogenide Monolayers}

\maketitle


\author{Pedro Miguel M. C. de Melo*}
\author{Zeila Zanolli*}
\author{Matthieu J. Verstraete}

\begin{affiliations}
Dr. Pedro Miguel Monteiro Campos de Melo
\liege\\ \utrecht \\\etsf\\
Email: p.m.monteirocamposdemelo@uu.nl\\
Prof. Dr. Zeila Zanolli\\
\barcelona\\ \utrecht\\ \etsf\\
Email: z.zanolli@uu.nl 
Prof. Dr. Matthieu Jean Verstraete\\
\liege\\ \etsf\\

\end{affiliations}



\begin{abstract}
Even the best quality 2D materials have non-negligible concentrations of vacancies and impurities. It is critical to understand and quantify how defects change intrinsic properties, and use this knowledge to generate functionality. This challenge can be addressed by employing many-body perturbation theory to obtain the optical absorption spectra of defected transition metal dichalcogenides. Herein metal vacancies, which are largely unreported, show a larger set of polarized exitons than chalcogenide vacancies, introducing localized excitons in the sub-optical-gap region, whose wave functions and spectra make them good candidates as quantum emitters. Despite the strong interaction with substitutional defects, the spin texture and pristine exciton energies are preserved, enabling grafting and patterning in optical detectors, as the full optical-gap region remains available. A redistribution of excitonic weight between the A and B excitons is visible in both cases and may allow the quantification of the defect concentration. This work establishes excitonic signatures to characterize defects in 2D materials and highlights vacancies as qubit candidates for quantum computing.
\end{abstract}

Transition metal dichalcogenides (TMDs) have become strong contenders for the engineering of optical devices, especially thanks to their coupling of the spin and valley degrees of freedom, and the presence of strongly bound excitons, opening avenues for next generation opto-electronics\cite{NatNano.7.699,Mak2016}. Manufactured samples have strongly improved in quality, but will always contain a significant concentration of defects\cite{Edelberg2019, Briggs_2019}. Graphene can be crystallized almost perfectly in very large flakes, but TMDs present many more natural defects, in particular chalcogen vacancies (see e.g. Ref.~\cite{PhysRevLett.109.035503}). On the bright side, defects and substitutional dopants can be used to tune the electronic structure and optical properties of materials~\cite{C6NH00192K}. By doing so, devices sensitive to specific wavelengths and polarizations can be engineered (reviewed in Ref.~\cite{Lin2016}), and can even behave as single photon emitters~\cite{Bourrellier2016}. There is an ongoing search for long lived spin states at room temperatures in TMDs. Here defects are expected to play a crucial role in both scattering and storing spin information - we showed recently that intrinsic scattering mechanisms (the electron-phonon interaction) can quickly destroy the pumped spin-polarisation~\cite{Ersfeld2019}. Chalcogen vacancies can also be used as grafting sites for functional groups, to create bio and chemical sensors\cite{Bolotsky2019,Azadmanjiri2020}. Alkane and other functional groups can be incorporated directly into the matrix (as opposed to thiol links, for instance, which necessitate Au).  Carbon atoms have also been used as acceptor dopants for bulk semiconductors~\cite{Yang1992}.  In TMDs, mixed phases with transition metal carbides have been shown to have applications in catalysis~\cite{Hai2017,Wu2017}, while in \mos{} it was shown that carbon substitutions had a strong effect on the TMD electronic and optical properties~\cite{Yue2013,PRB.95.245435}. 

Many experiments give access to the presence and properties of localized defects. In TMDs the most commonly used are: Scanning Tunnelling Microscopy which shows contrast changes due to chemical substitution and electron cloud reconstruction around defects\cite{Schuler2019}; Scanning Tunnelling Spectroscopy, which probes the detailed electronic structure at the defect site\cite{Schuler2019}; optical spectroscopy showing absorption and photoluminescence by the defect-induced states\cite{Carozo2017, Jeong2019, Dang2020}; tunnelling transport from insulated contacts through a core material, which is resonant through defect states in the core band gap\cite{Jeong2019}; transmission electron microscopy, which gives both structural and chemical information\cite{Edelberg2019, Hong2020}.

Understanding how defects affect optical properties is a first essential step towards controlled functionalization of materials, both for fingerprinting (optical characterization is simple, remote, and non destructive) and to understand derived optical functionalities. First-principles computational techniques provide a high degree of physical insight and predictive power in the spectral features of defects, to find new peaks and yield quantitative positions and weight transfers. 

In this work, we present a fully First-principles investigation of defected monolayers of \ws{}, based on the Bethe Salpeter Equation (BSE) for electron-hole interactions within many body perturbation theory. We analyze the resulting changes in electronic band structure and optical absorption spectra, aiming to answer the question: can we identify a defect, and ideally quantify its concentration~\cite{Refaely-Abramson2018}, just by looking at the absorption spectrum? While chalcogenide vacancies have been subject to some studies\cite{Refaely-Abramson2018}, metal vacancies and isovalent substitutions like \mow{} and \ch{} are still largely unreported, limiting our knowledge of their behaviour as quantum emitters or chemical detectors.
We find that defects fall into two functional categories, based on the presence of bound states within the band-gap of \ws{}. We discuss their spectra, spin textures and the criteria which could be used to identify each defect. 

\begin{figure}[htp]
    \includegraphics[width=0.9\columnwidth]{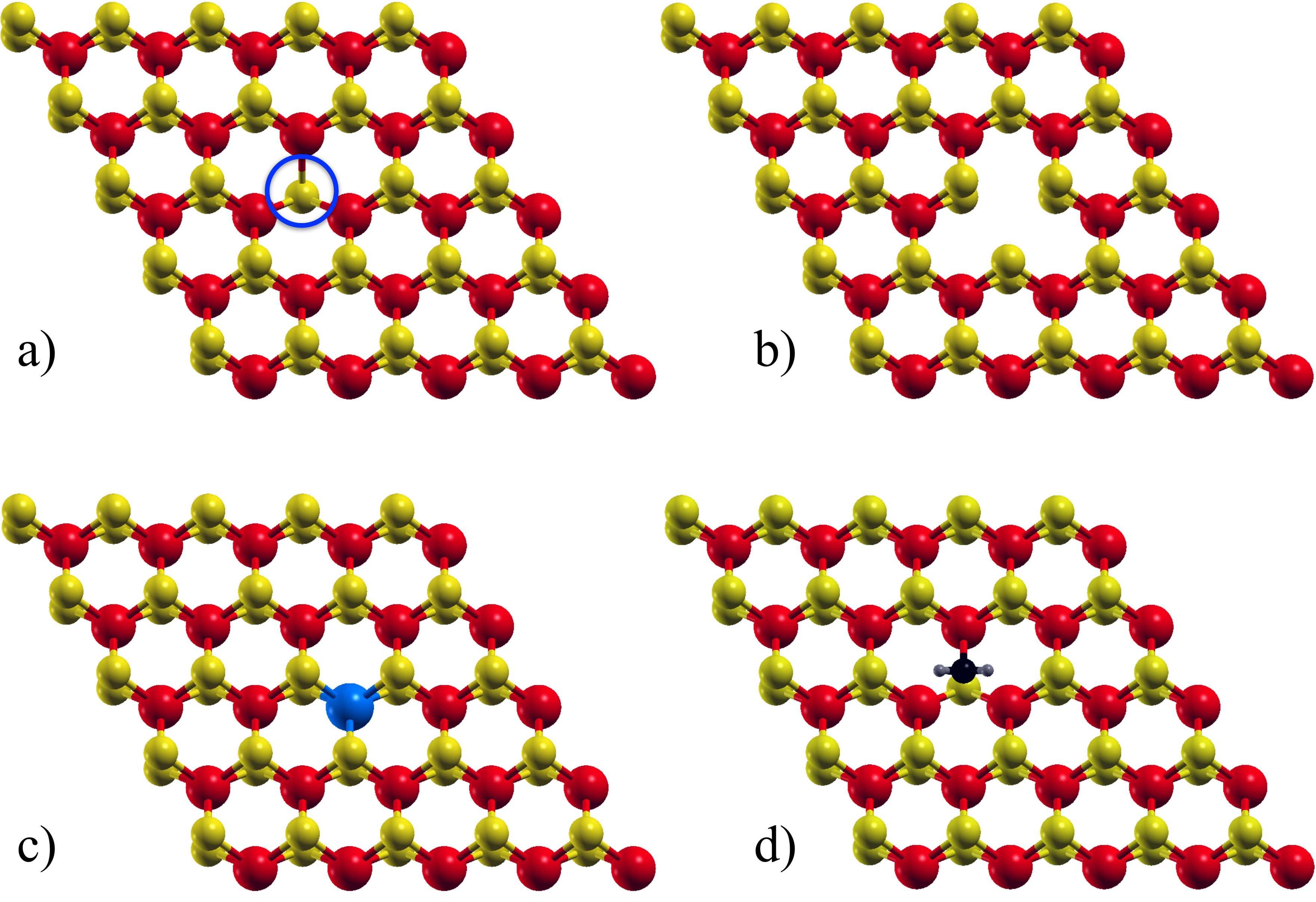}
    \caption{Defected \fbf supercells used in this work. a) S vacancy, marked by the blue circle. b) W vacancy. c) \mow{} substitution. d) \ch{} substitution. Tungsten atoms are shown in red, sulfur in yellow, molybdenum in blue, carbon in black, and hydrogen in grey.}
    \label{fig2}
\end{figure}

In \textbf{Figure~\ref{fig2}} we show the four defects that are the focus of our work: two vacancies (S and W ions); and two substitutions, \mow{} and  \ch{}. The S vacancy is the most commonly found defect in monolayers of \ws{} and often assigned to specific features below the optical gap~\cite{Carozo2017}. In the substitution case \mow{} is quite commonly found in nature, and carbon is a common dopant in semiconductors~\cite{acsnano.9b02316}. Studies have been made on the potential transport applications of TMDs and transition metal carbides~\cite{Hai2017,Jeon2018}. In the case of \mos{} experiments point to changes to the electronic structure due to carbon doping, which should translate into new optical features~\cite{Yue2013,Hu2015}.

In ideal defect engineering, induced changes in the system should yield new controllable features that are distinct from the pristine properties. In the case of TMDs, a key property that relates to optics is the polarisation of states at the K wave vector in the Brillouin zone. The polarization will control the allowed optical transitions that form bright excitons. We label in order of increasing energy the last two occupied states and the first two unoccupied states at K as $(v_1,v_2,c_1,c_2)$. For pristine \ws{} (see the Supplementary Material for more information), the first bright exciton would be made of an optical transition from $v_2$ to $c_1$, while the second brightest exciton would be made by a transition from $v_1$ to $c_2$. This is due to the fact that optical selection rules enforce spin conservation in TMDs: the valence and conduction manifold differ in orbital character, guaranteeing the needed angular momentum change when absorbing a helical photon~\cite{Molina-Sanchez2017}. 

We start by analyzing two systems which feature states in the pristine band gap: the sulfur and  tungsten vacancies. Their DFT band structures are shown in Figure~\ref{fig2} a) and~\ref{fig2} c), respectively. Note that using a \fbf supercell, the point K of the pristine Brillouin zone (BZ) folds back into the point K of the supercell BZ.

Both systems preserve the spin polarisation of pristine states at K, meaning that the A and B peaks that are the dominant features of the pristine spectrum must also be present. For the S vacancy, Figure~\ref{fig2} a), new non-dispersive mid-gap singlet states appear 1.08 and 1.28 eV above the valence band edge, in agreement with what was previously found~\cite{Schuler2019a,PRB.95.245435}. At K the first band is almost completely spin-unpolarized, while the second is completely polarized. This opens a path for new emission channels, as electronic transitions from the pristine valence states to these new defect states are allowed, with different oscillator strengths and different excitation energies.

The W vacancy is more challenging numerically: the defect bands are split off from the bulk conduction and valence bands, but in the \fbf supercell they still show some dispersion and a finite band width. Their position and spin texture are close to what has been found in other DFT calculations for this defect using larger supercells~\cite{PRB.95.245435}, and we extract a semi-quantitative picture of the optical properties. Together with the six new mid-gap states, there are also four new occupied states bound to the defect in the valence region (unlike the S vacancy). All these states are spin polarized at K, so if we consider optical selection rules, we can expect new peaks to show up in the absorption spectrum. These peaks will be a combination of transitions between pristine and defect states, and others between defect states (occupied to empty) in the mid-gap region.

The effects of changes in the band structure on the BSE optical spectrum are shown in \textbf{Figure~\ref{fig3}}. For comparison the absorption spectrum of the pristine system is shown in grey. The energies of all identified excitons are listed in Table~1 in the SI, along with the reference energies for the pristine case. Note in passing that the higher energy peaks beyond A and B are not reproduced in full in our calculations, due to limitations in the number of states included in the BSE for such large systems.

With an S vacancy in Figure~\ref{fig2} a) two new peaks arise at 1.35 and 1.14 eV due to the mid-gap states. They are marked D$_1$ and D$_2$ in Figure~\ref{fig2} b). As shown in the inset, \dd{1} and \dd{2} correspond to optical transitions from the top valence band to the defect state at 1.28 eV and 1.08 eV, respectively. However, since the lowest defect state is not fully polarized, the resulting dipole matrix element is much smaller than that of the \dd{1} peak, where the defect state is fully polarized. 

The exciton wave function for the peak corresponding to \dd{1} is shown in Figure~\ref{fig3} a). Here the hole is placed at the position of the vacant sulfur ion (marked by the green sphere) and the magnitude on the colour-map shows the probability density of the electron (see the Figure 5 in the SI where all the excitonic wave functions are depicted). The color map shows that exciton states are highly localized on the neighboring tungsten atoms, indicating that the S vacancy does indeed form a quantum dot.  

For the W vacancy, whose spectrum is shown in Figure~\ref{fig2} d), more peaks are present due to the increased number of defect states. In total we identify five excitons: three composed from holes in the valence band and electrons in defect states, \dd{1} to \dd{3}; and two others with both the electron and the hole bound to two different manifolds of defect states, \dd{4} and \dd{5}. The states involved in the formation in the excitons are shown in the inset of Fig~\ref{fig2} d). The most striking feature is the relative intensity of the \dd{4} and \dd{5} excitons relative to the A and B excitons. As defect states have very weak dispersion, the associated electrons and holes will have large effective masses. It is possible to show~\cite{Hamaguchi2017} that oscillator strengths are proportional to the reduced mass of electron and hole, which explains why the effect is further magnified for transitions \emph{between} defect states in the case of the W vacancy.

A more detailed analysis to the excitonic wave functions shows that two of the low intensity peaks (\dd{1} and \dd{3}) are artefacts due to the interaction of defect states in neighboring periodic replicas. They result from electron-hole pairs localized on different vacancy sites (see Figure 7 in the SI for zoomed out plots of excitonic wave function). 
The majority of the electronic charge in the exciton is not localized near the same ion vacancy as the hole, and is bound thanks to a finite overlap of the DFT defect state wavefunctions. These two peaks will disappear if larger supercells are used: the dipole matrix element and oscillator strength will go do 0. We note that while the BSE is solved only for $\qq = 0$, the introduction of non-dispersive defect states allows for many transitions to occur throughout the BZ, so an exciton function can actually be composed of several vertical transitions at different momenta (NB: this is distinct from a finite wave vector for the whole exciton).  

In the case of \dd{2}, shown in Figure 7 in the SI, there is a residual interaction with adjacent vacancy sites, but now the hole is correctly bound to an electron located at the same site. For \dd{5}, shown in Figure~\ref{fig3} b) both the electron and hole are entirely located at the same site. 

We studied two cases of substitutions in the \ws{} monolayer; one with a molybdenum atom replacing a tungsten atom; and another where a sulfur atom was replaced by a methyl, the simplest (divalent) group representing grafted organic substituents.  

In both cases no mid gap states where found (see Figure 4 in the SI for their bandstructures), which can be rationalized as follows. In the case of \mow{} substitution, molybdenum and tungsten have the same valence, close atomic and covalent radii\cite{B801115J,doi:10.1002/chem.200901472,doi:10.1002/chem.200800987}, resulting in similar chemical properties. A small concentration of defects does not lead to strong changes in the charge density, thus leaving the system practically unchanged when compared to the pristine case. For the \ch{} substitution, the methylene group provides the same number of valence electrons as the sulfur atom. The breaking of local symmetry is not strong enough to perturb the band edges.

The lack of mid-gap states is reflected in the absorption spectra shown in \textbf{Figure~\ref{fig4}}. In both cases the excitonic peaks lie almost on top of those of the pristine system, with the largest deviation being 60 meV for the \mow{} A peak. The energies for the defected A and B peaks are shown in Table 1 in the SI.

There is, however, both weight transfer and changes in spin texture for the A and B peaks at higher energies in the absorption spectrum of both substitutions, which suggests methods to identify these defects experimentally. The strongest signature of the substitution lies within the exciton wave functions shown in \textbf{Figure~\ref{fig5}}. While the exciton cloud is still dispersed throughout the crystal, there is a higher charge concentration near the substitution. In the \ch{} case this is valid for the A exciton (Figure~\ref{fig5} c)), and in \mow{} both A and B excitonic states localize near the defect. The \ch{} case also shows breaking of $C_3$ symmetry by the methyl molecule. The localization is due both to the defect-related electronic states and to the choice of the initial position of the hole. For reference, in boron nitride similar extensions of 3-5 nearest neighbors are found in Ref.~\cite{Paleari2018}.
   
The two peaks which correspond to the bulk excitons A and B known from literature, and are within 50 meV of those of the pristine system in all cases (this is below the absolute precision of the first principles methods, and shows the basic convergence of our supercell sizes).

We can now establish a more complete picture of how different defects change optical properties of TMDs. The main changes to the optical spectrum come from new mid-gap states. Isovalent substitutions like \mow{} and \ch{} will not be trivially seen in the absorption spectra, but can still be detected by the ratio of the A and B peak intensities, and by localisation in the exciton's spatial distribution. 

In terms of potential applications, the two vacancies are clear front runners for designing quantum dots and quantum emitters. In particular the S vacancy has already shown some promising results as a single photon emitter~\cite{He2015}. The two mid-gap defect states are actively considered for quantum computing applications: they are separated in energy by 0.21 eV, making them addressable using mid-infrared lasers, and insulating them from the highest phonon energy in pristine \ws{}, which is 53 meV~\cite{Pike2018,Pike2019}. 

The W vacancy offers a larger set of localized excitons, but is more energetic and harder to produce (see Table 1 in the SI). Here the brightest excitons are made of transitions between single particle defect states, due to their large effective masses. These exciton states show potential in devices as they would behave as bright emitters with multiple internal states, behaving like an embedded molecule for multivalued quantum computing\cite{Fresch2018}. The residual band width will disappear only in a 7\by7 supercell\cite{PRB.95.245435}, but the spin texture and qualitative features are already well represented here (see Figure 3 in the SI). In actual samples, this vacancy type is more likely to be charged, with electrons filling the dangling S ion's orbitals.

We have shown that the isovalent substitutions \mow{} and \ch{} will not produce in-gap states. Though this is intuitive, it is not trivial, and the local electronic structure is strongly modified as shown by the excitonic wave functions. For purposes of grafting organic molecules, this means the full band gap window of intrinsic \ws{} will be available for optical sensing.

In conclusion, we employ powerful and accurate first principles techniques to shed some light into the changes in optical properties introduced by point defects in TMDs. Two promising systems, S and W ion vacancies, have potential for quantum emitters and even quantum computing. Isovalent vacancies do not introduce mid-gap states in the band structure, but do change the relative intensity of the canonical A and B peaks, and the spatial distribution of the first and second excitonic states, in the \ch{} case even leading to a breaking of symmetry. In both cases the full sub optical gap region is available for use in detection of molecules that might graft themselves onto the TMD's surface.

\medskip
\textbf{Supporting Information} \par Supporting Information is available from the Wiley Online Library. Additional details on the convergence with cell size, the wave functions for all defect-bound exciton states, and DFT and Many-body Perturbation theory methods and numerical parameters used in our calculations.

\medskip
\textbf{Acknowledgements} \par
We wish to acknowledge important input, discussions, and stimulus from M. Terrones, B. Biel, and M. Palummo, as well as extensive support from the \yambo developer team.

PMMCM and MJV acknowledge funding by the Belgian FNRS (PDR G.A. T.1077.15, T.0103.19, and an ``out'' sabbatical grant to ICN2 Barcelona), and the Communaut\'e Fran\c{c}aise de Belgique (ARC AIMED G.A. 15/19-09). This publication is based upon work of the MELODICA project, funded by the EU FLAG-ERA\_JTC2017 call.

The work benefited from HPC-EUROPA3 (INFRAIA-2016-1-730897) H2020 Research Innovation Action hosted by the Theory and Simulation group at ICN2 supported by the Barcelona Supercomputing Center, and from the access provided by ICN2 (Barcelona, Spain) within the framework of the NFFA-Europe Transnational Access Activity (grant agreement No 654360, proposal ID 717, submitted by PMMCM). 

Z.Z. acknowledges support by the Ram\'on y Cajal program RYC-2016-19344 (MINECO/AEI/FSE, UE), Spanish MINECO (FIS2015-64886-C5-3-P), the Severo Ochoa Program (MINECO, SEV-2017-0706), the CERCA programme of the Generalitat de Catalunya (Grant 2017SGR1506), the EC H2020-INFRAEDI-2018-2020 MaX Materials Design at the Exascale CoE(grant No. 824143), and the Netherlands sector plan program 2019-2023.

Computational resources have been provided by the Consortium des Equipements de Calcul Intensif (CECI), funded by FRS-FNRS G.A. 2.5020.11; the Zenobe Tier-1 supercomputer funded by Walloon G.A. 1117545; and by PRACE DECI grants 2DSpin and Pylight on Beskow (G.A. 653838 of H2020, and FP7 RI-312763). The authors thankfully acknowledge the computer resources at Mare Nostrum technical support provided by the  Barcelona Supercomputing Center (Spanish Supercomputing Network, RES). This publication is based upon work from COST Action TUMIEE (CA17126), supported by COST (European Cooperation in Science and Technology).

\medskip
\textbf{Conflicts of Interest}\par
The authors have no commercial or financial conflicts of interest. 

\medskip
\bibliographystyle{MSP}
\bibliography{literature}

\begin{figure}[htp]
    \includegraphics[width=\textwidth]{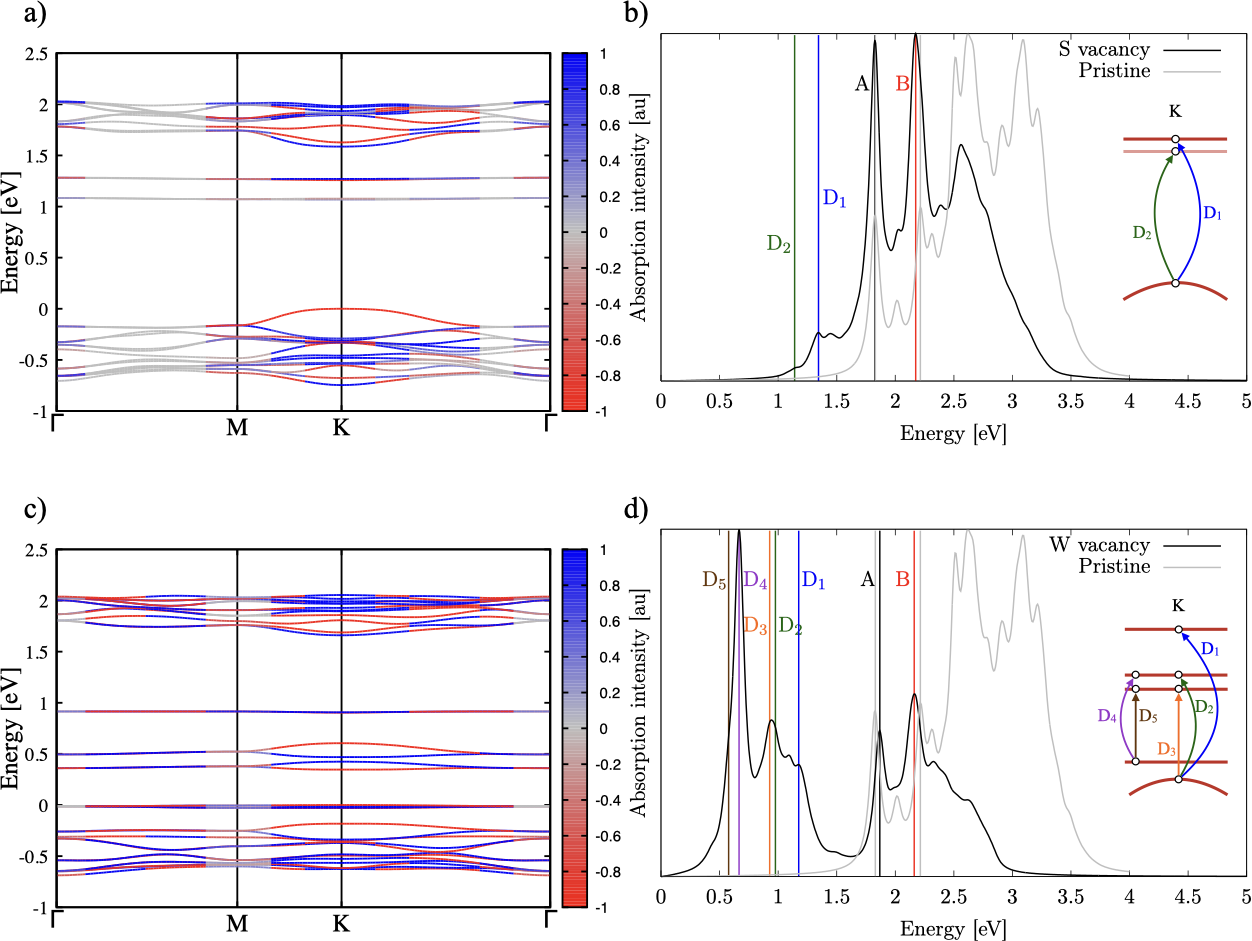}
    \caption{(color online) a) and c): DFT band structures and spin-texture of defected \ws{} for the S and W vacancies, respectively. Color scale indicates value of $\braket{S_z}$ (red for 1 and blue for -1). The reference Fermi level at 0 is set to the last occupied state. In both cases new states arise in the mid-gap region, with new occupied defect states also showing in the W vacancy's band structure. b) and d): Optical absorption spectra of the S and W vacancies, respectively. Absorption of the pristine system is shown in grey. The positions of the first excitonic peaks are shown by vertical lines. Labels and vertical lines have matching colours. Two new peaks arise for the S vacancy due to the two new manifolds of mid-gap states. The W vacancy exhibits several new peaks, as four new manifolds of defect states arise. The insets in b and d show the level scheme and main excitonic transitions.}
    \label{fig1}
\end{figure}

\begin{figure}[tp]
    \includegraphics[width=0.5\columnwidth]{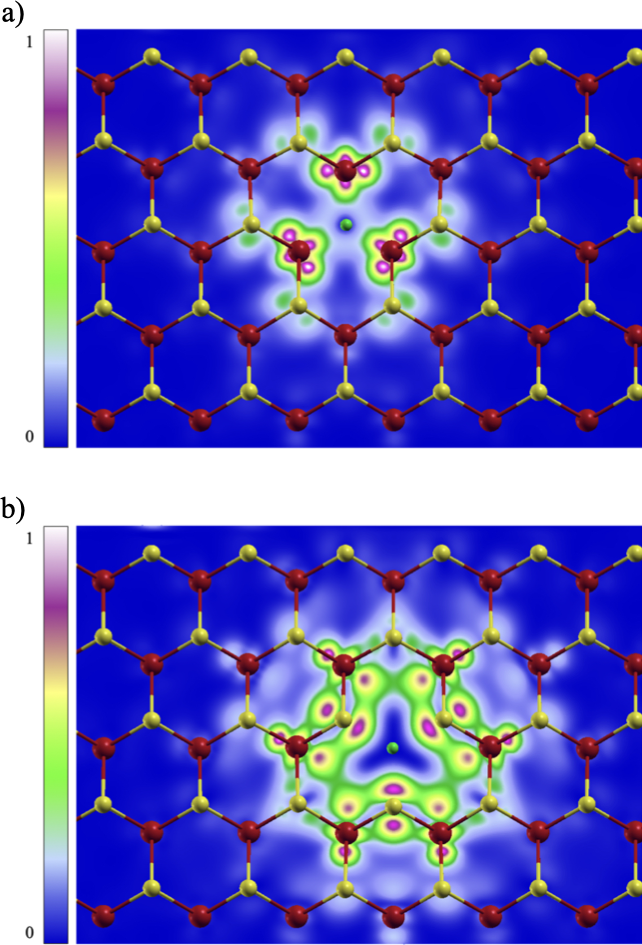}
    \caption{(color online) a) S vacancy wave function for excitonic state \dd{2}, shown in Figure~\ref{fig2} b). b) W vacancy wave function for excitonic state \dd{5}, shown in Figure~\ref{fig2} d). The hole is marked by the green sphere and it is placed at the position of the missing sulfur ion. In both cases the electronic charge is highly localized near the hole, indicating the formation of a quantum dot.}
    \label{fig3}
\end{figure}

\begin{figure}[htp]
    \includegraphics[width=0.5\columnwidth]{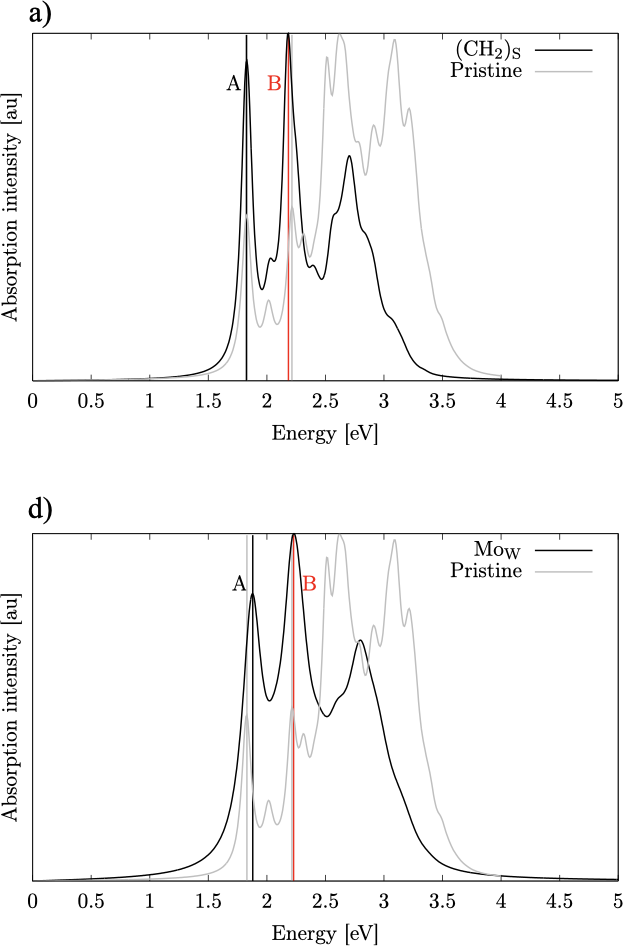}
    \caption{(color online) Optical absorption spectra for the 5$\times$5 supercells with the a) Mo$_\mathrm{W}$ substitution; b) \ch{} substitution. Absorption of the pristine system is shown in grey. The positions of the first excitonic peaks are shown by vertical lines. Labels and vertical lines have matching colours. Both cases show no new peaks, with the A and B excitons at energies that almost match those of the pristine case.}
    \label{fig4}
\end{figure}

\begin{figure}[htp]  
    \includegraphics[width=\columnwidth]{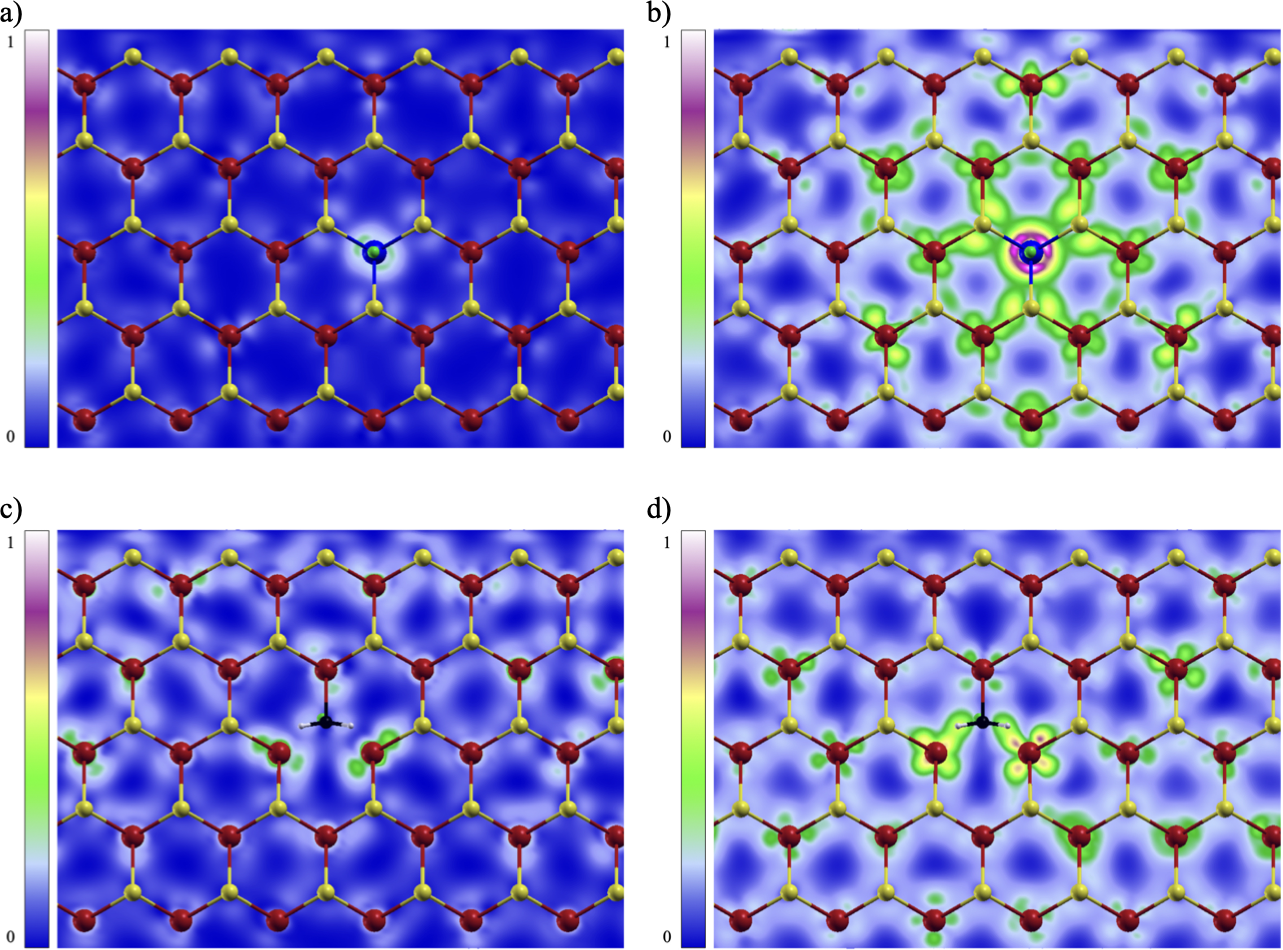}
    \caption{A (left column) and B (right) excitonic wave functions of \mow{} (a) and b)) and \ch{} (c) and d)), shown in Figure~\ref{fig4}. The hole is marked by the green sphere and it is placed between the the mid-plane and the carbon ion. While there is some charge localisation and symmetry breaking, the exciton wave function extends through the whole system.}
    \label{fig5}
\end{figure}


\begin{figure}
\textbf{Table of Contents}\\
\medskip
  \includegraphics{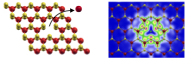}
  \medskip
  \caption*{Large and defectless monolayers of transition metal dichalcogenides (TMDs) are difficult to manufacture, with most of them containing defects. Herein it is shown using Many-Body Perturbation Theory how different vacancies and substitutions can in fact be advantageous, by creating new quantum dots or adsorption sites in the monolayer, thus increasing it functionalization potential.}
\end{figure}

\end{document}


\pagestyle{fancy}
\rhead{\includegraphics[width=2.5cm]{vch-logo.png}}

\title{Supplementary Material for: Optical Signatures of Defect Centres in Transition Metal Dichalcogenide Monolayers}
\maketitle

\author{Pedro Miguel M. C. de Melo*}
\author{Zeila Zanolli*}
\author{Matthieu J. Verstraete}

\begin{affiliations}
Dr. Pedro Miguel Monteiro Campos de Melo
\liege\\ \utrecht \\\etsf\\
Email: p.m.monteirocamposdemelo@uu.nl\\
Prof. Dr. Zeila Zanolli\\
\barcelona\\ \utrecht\\ \etsf\\
Email: z.zanolli@uu.nl 
Prof. Dr. Matthieu Jean Verstraete\\
\liege\\ \etsf\\

\end{affiliations}

\date{\today}

\begin{abstract}
In the Supplementary Information we present more details about our calculations, including numerical parameters used in our DFT and Many-Body theory calculations. We also discuss the convergence of with respect to supercell size and how 3\by3 supercells are insufficient to ensure non-dispersive defect states. Finally we show the wave-functions of all exciton states associated with defect levels in the w vacancy case to understand which ones should vanish when using larger cell sizes.

\end{abstract}

\section{Pristine system}
\begin{figure}[htp]
    \begin{subfigure}[b]{0.49\textwidth}
    \includegraphics[width=\textwidth]{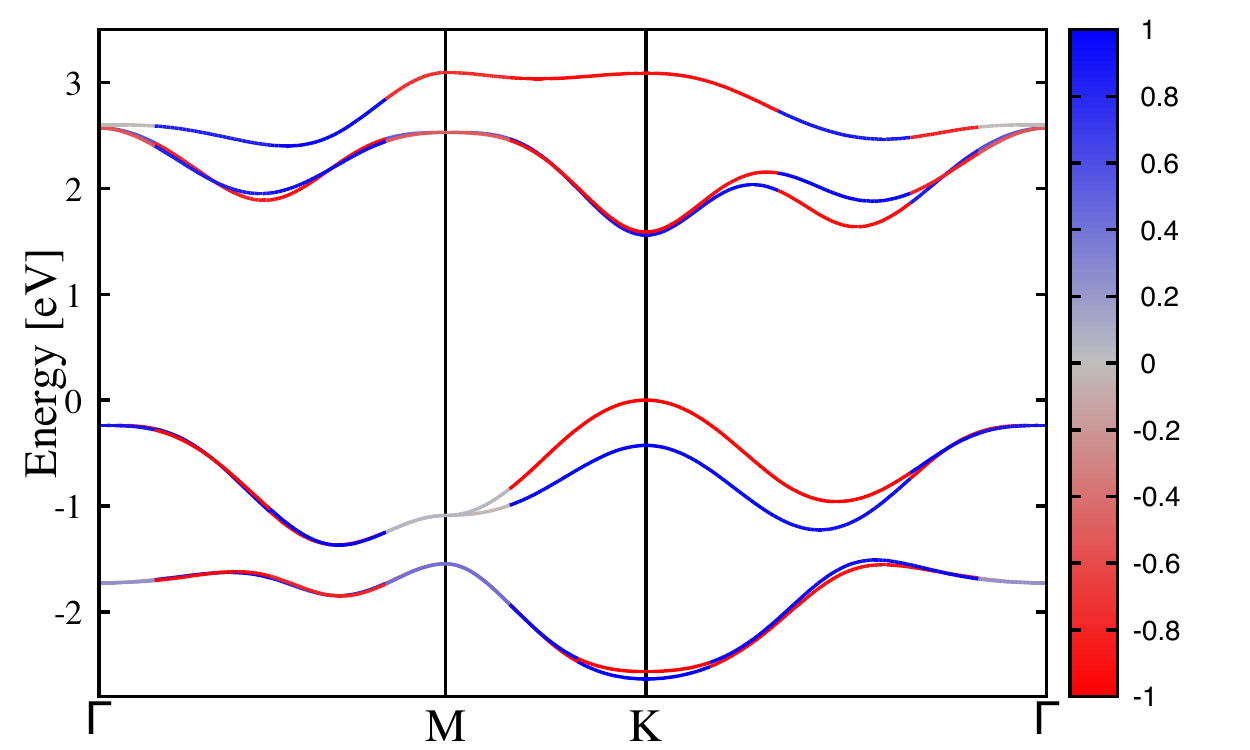}
    \caption{Bands structure of the pristine \ws{} monolayer.}
    \label{bands}
    \end{subfigure}
    \begin{subfigure}[b]{0.49\textwidth}
    \includegraphics[width=\textwidth]{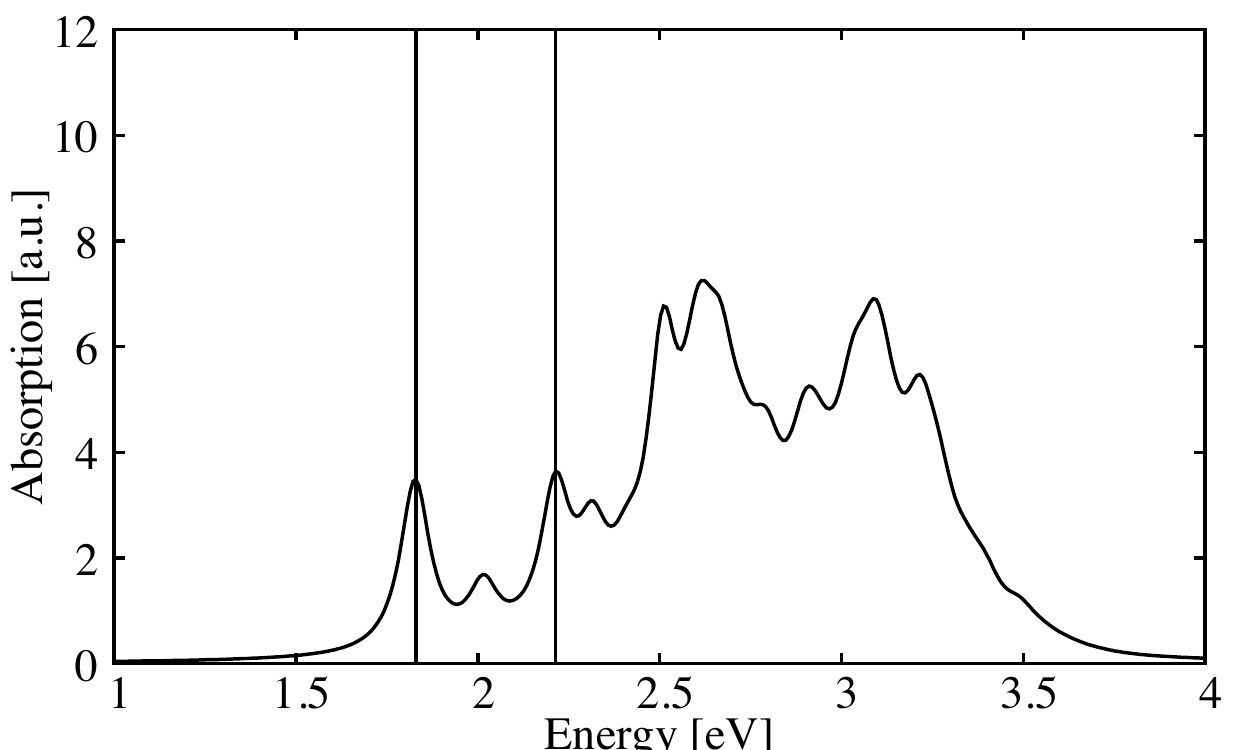}
    \caption{BSE absorption spectrum of a pristine \ws{} monolayer.}
    \label{bse}
    \end{subfigure}
    \caption{(color online) a) DFT band structure for pristine \ws{} (reference system). color scale indicates value of $\braket{S_z}$ (yellow for 1 and black for -1). b) BSE absorption spectrum and vertical lines indicating the positions of the two first bright excitons (A and B).}
    \label{fig5}
\end{figure}

The building block of our calculations was unit cell of pristine \ws{}. The lattice constants were set to 3.187 \AA and 30 \AA. The ground state was computed on a 24$\times$24 Monkhorst-Pack k-point grid, which was then used to generate the full set of occupied and unoccupied states on a 36\by 36 k-point grid. An energy cut-off of 120 Ry was used for the wave-functions. The resulting band structure and spin texture are shown in Fig.~\ref{bands}.  

Figure~\ref{bse} shows the resulting optical spectrum obtained from a \gw and BSE calculation, with the A and B exciton energies (1.83 and 2.21 eV, respectively) marked by vertical lines. In those calculations the G-vector's cut-off was set at 3500 meV. These parameters were later checked for convergence in GW and BSE calculations and were found to be enough to ensure converged results.

\section{Supercell size convergence and structural changes}

\begin{figure}[htpb]
    \begin{subfigure}[b]{0.49\textwidth}
        \includegraphics[width=\textwidth]{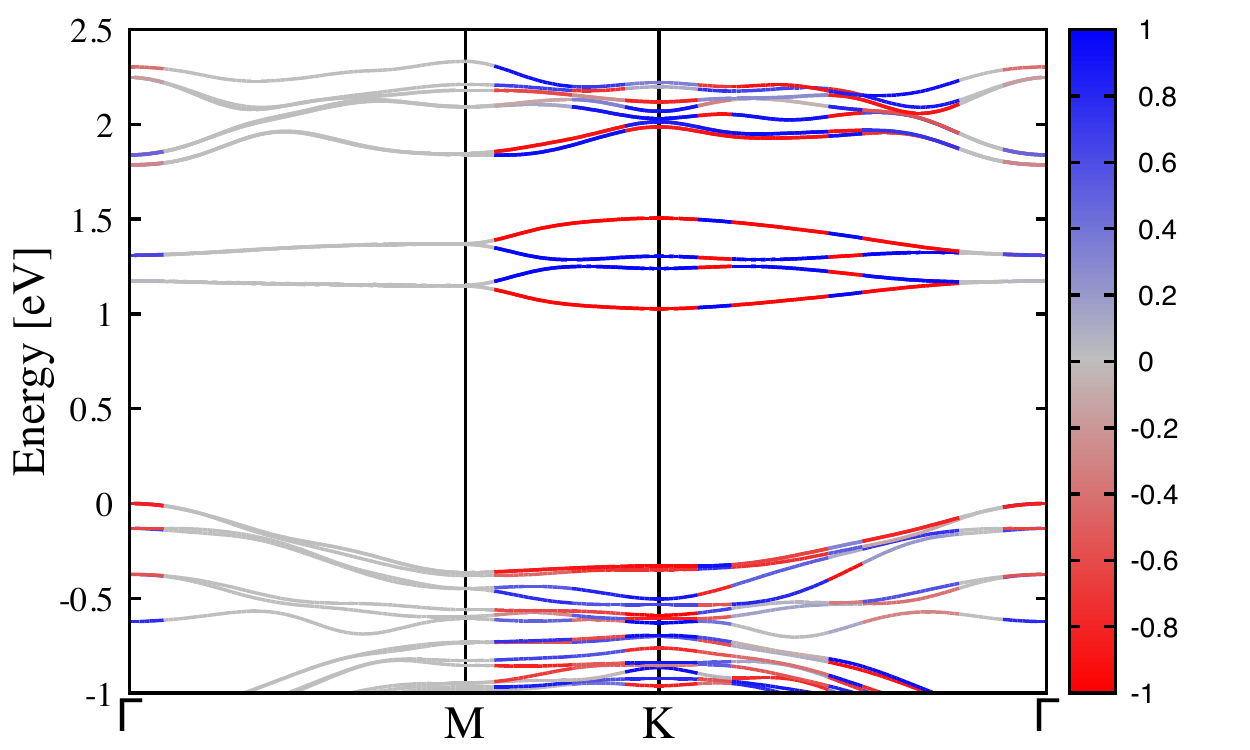}
        \caption{S vacancy.}
        \label{sv33bands}
    \end{subfigure}
    \begin{subfigure}[b]{0.49\textwidth}
        \includegraphics[width=\textwidth]{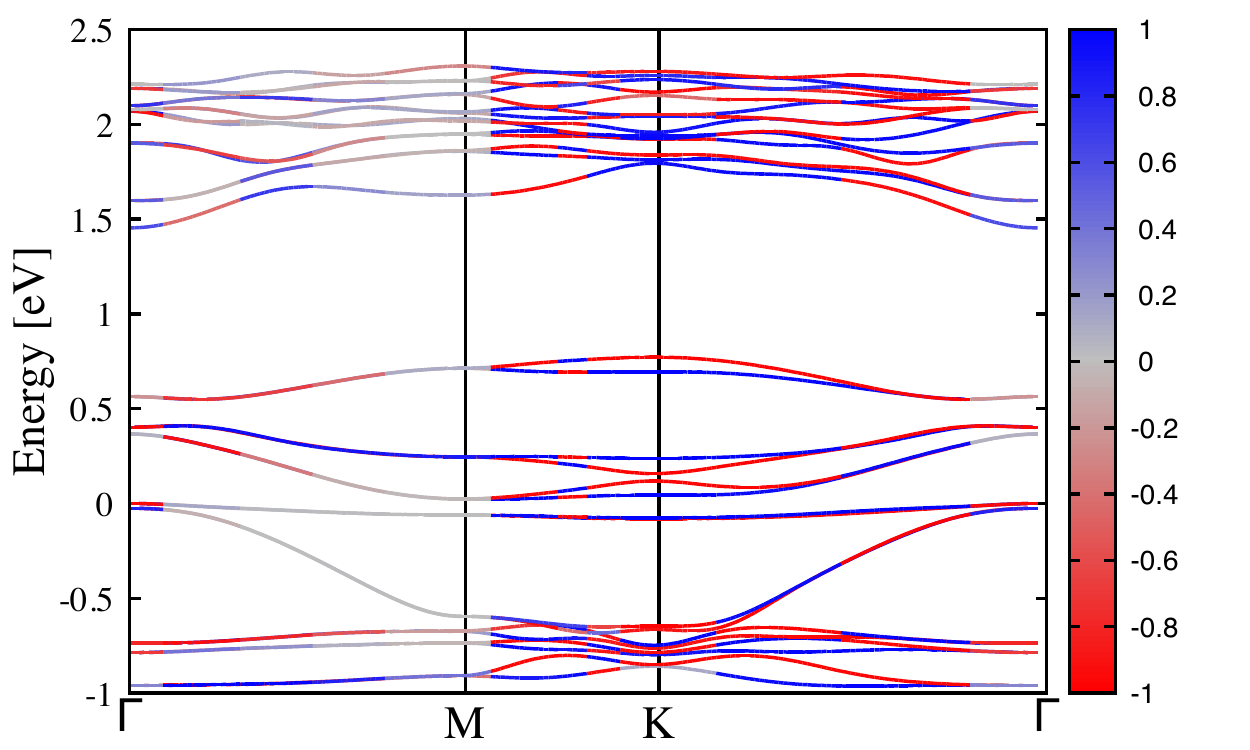}
        \caption{W vacancy.}
        \label{wv33bands}
    \end{subfigure}
    \begin{subfigure}[b]{0.49\textwidth}
        \includegraphics[width=\textwidth]{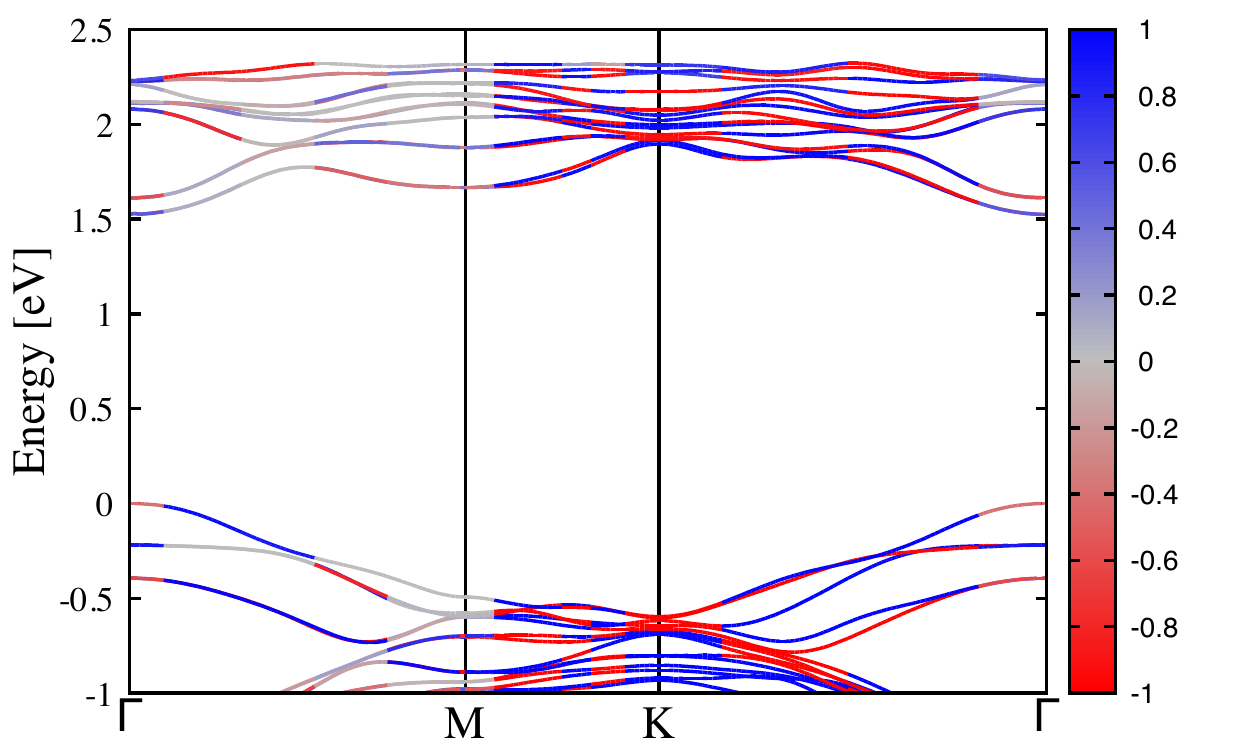}
        \caption{\mow{} substitution.}
        \label{mor33bands}
    \end{subfigure}
    \begin{subfigure}[b]{0.49\textwidth}
        \includegraphics[width=\textwidth]{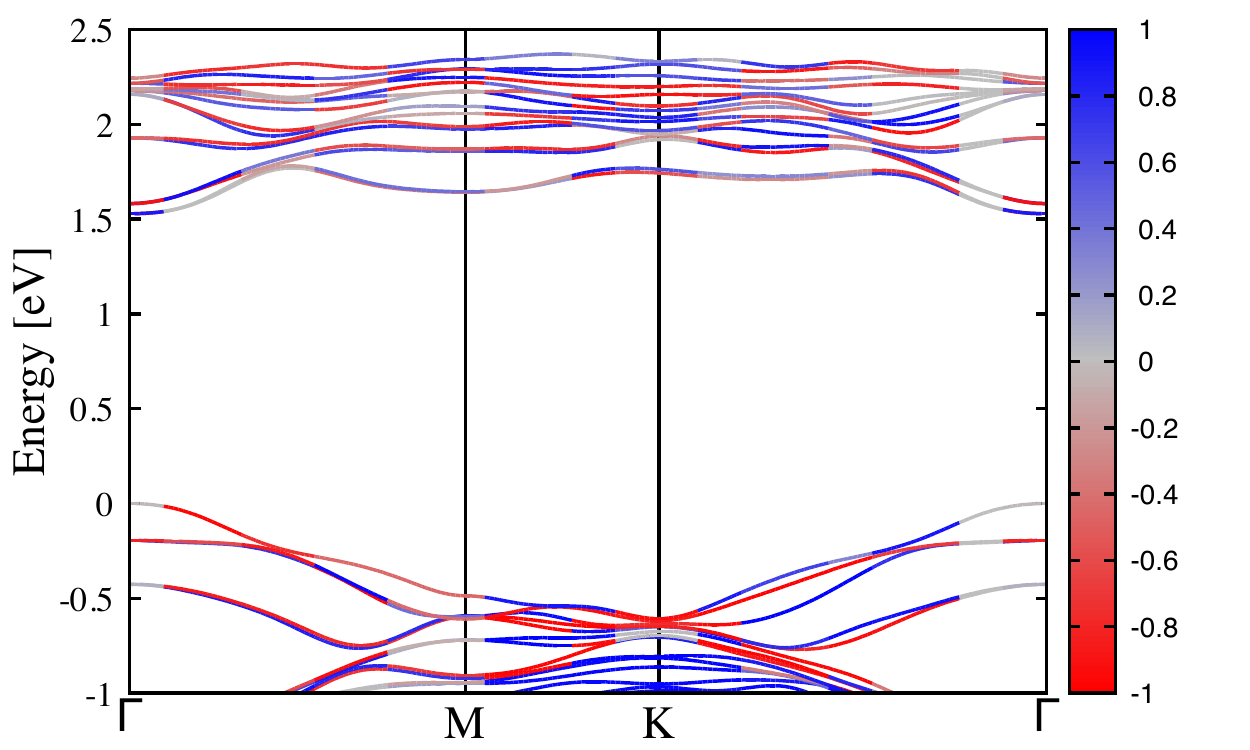}
        \caption{\ch{} substitution.}
        \label{ch233bands}
    \end{subfigure}
    \caption{(color online) DFT band structures and spin-texture of defected \ws{} on 3\by 3 supercells. a) S vacancy; b) W vacancy; c) Mo$_\mathrm{W}$ substitution; d) \ch{} substitution. The color scale indicates value of $\braket{S_z}$ (red for -1 and blue for 1). The presence of dispersive bands in (a) and (b) indicates that the supercells are too small, and so defect sites will interact with each other.}
    \label{fig7}
\end{figure}

Our 3\by 3 and 5\by 5 supercells were generated by replicating the relaxed pristine cell and then introducing the defect. Atomic coordinates were then relaxed keeping the lattice parameter fixed, starting from the 3\by 3 supercells. To understand if the cell size was enough to ensure meaningful results, we looked first at the band-structures, shown in Fig.~\ref{fig7}. 

\begin{figure}[htpb]
    \includegraphics[width=\textwidth]{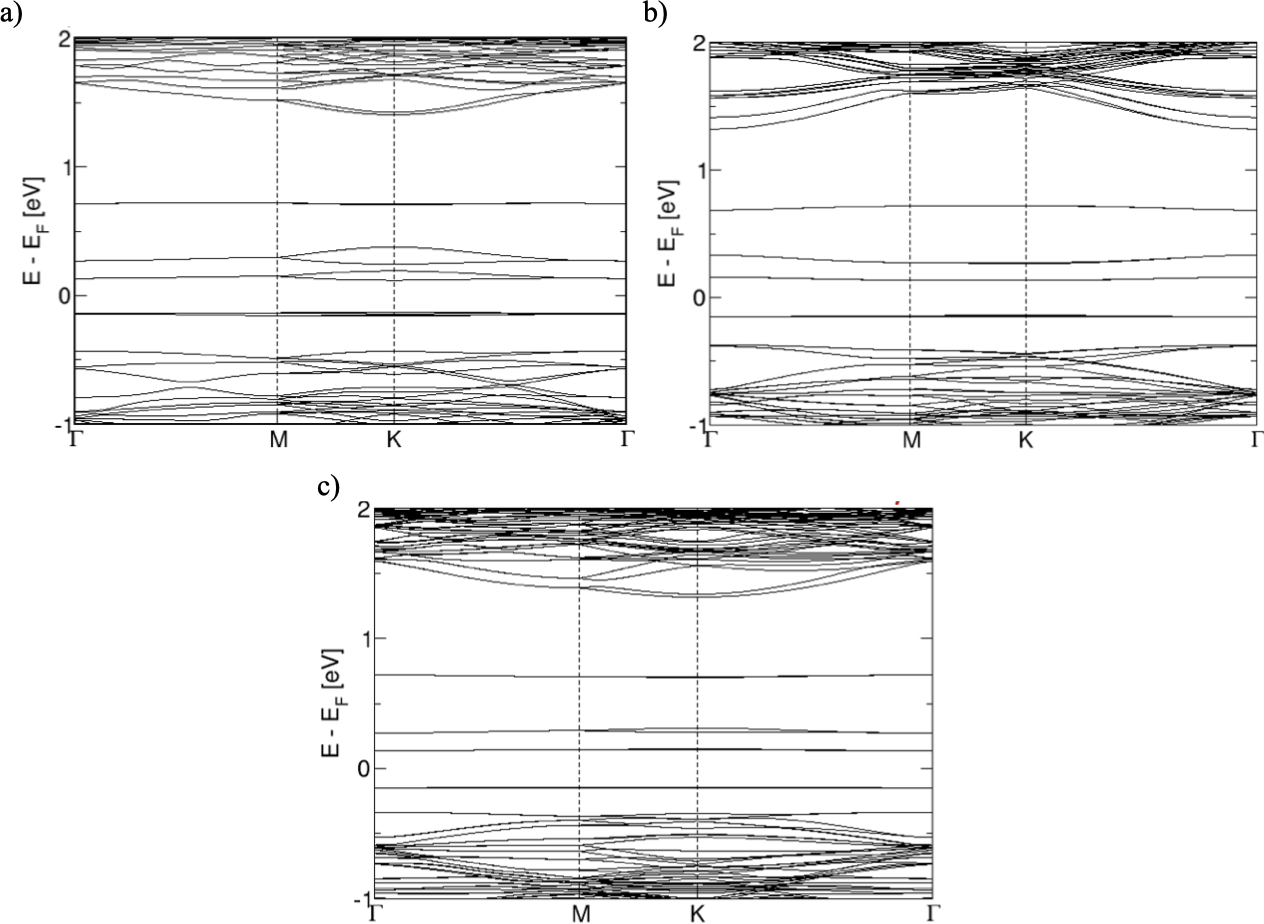}
    \caption{(color online) DFT band structures and spin-texture of a \ws{} monolayer with a W vacancy on a 5\by5 (a)), 6\by6 (b)), and 7\by7 (c)) supercell obtained with the SIESTA simualtion package~\cite{Soler_2002}. The 5\by5 results are in good agreement with those from Quantum Espresso, while those from the 7\by7 supercell show the need to use larger cell sizes when dealing with metallic vacancies in TMDs.}
    \label{fig9}
\end{figure}

Both vacancies show new mid-gap states, but these are still dispersive. For the S vacancy, Fig.~\ref{sv33bands}, the new states are already at close enough in energy to the positions we obtain for the 5\by5 supercell, but remain dispersive and non-degenerate near K. The W vacancy, Fig.~\ref{wv33bands}, is still far from convergence, with mid-gap states being so close to the top of the valence band that the system can easily turn metallic. As shown in Fig.
~\ref{fig9}, a 7\by7 supercell is the smallest size for a TMD supercell with a metallic defect. For both \mow{} and \ch{}, shown in Figs.~\ref{mor33bands} and~\ref{ch233bands}, respectively, the band-structures showed already that no mid-gap states were present, indicating that high concentrations of these defects will have very little impact in absorption properties. 

\begin{figure}[htpb]
    \begin{subfigure}[b]{0.49\textwidth}
        \includegraphics[width=\textwidth]{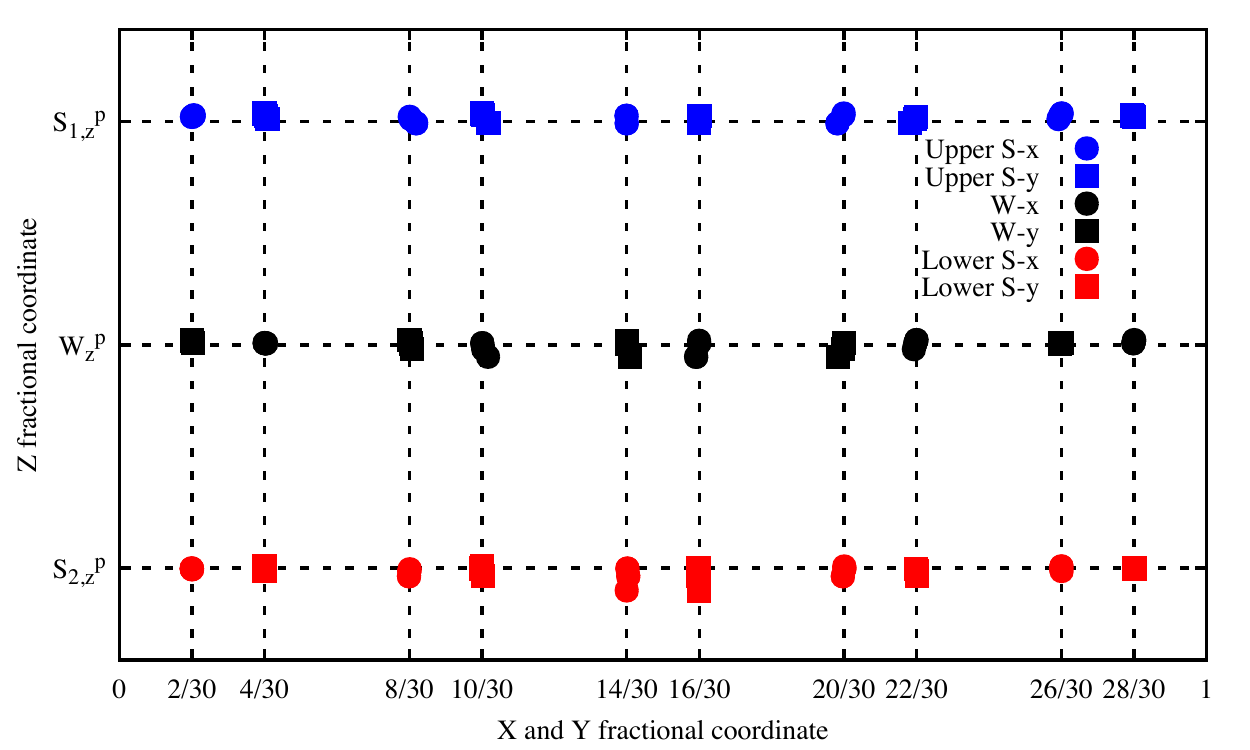}
        \caption{S vacancy.}
        \label{sv55coords}
    \end{subfigure}
        \begin{subfigure}[b]{0.49\textwidth}
        \includegraphics[width=\textwidth]{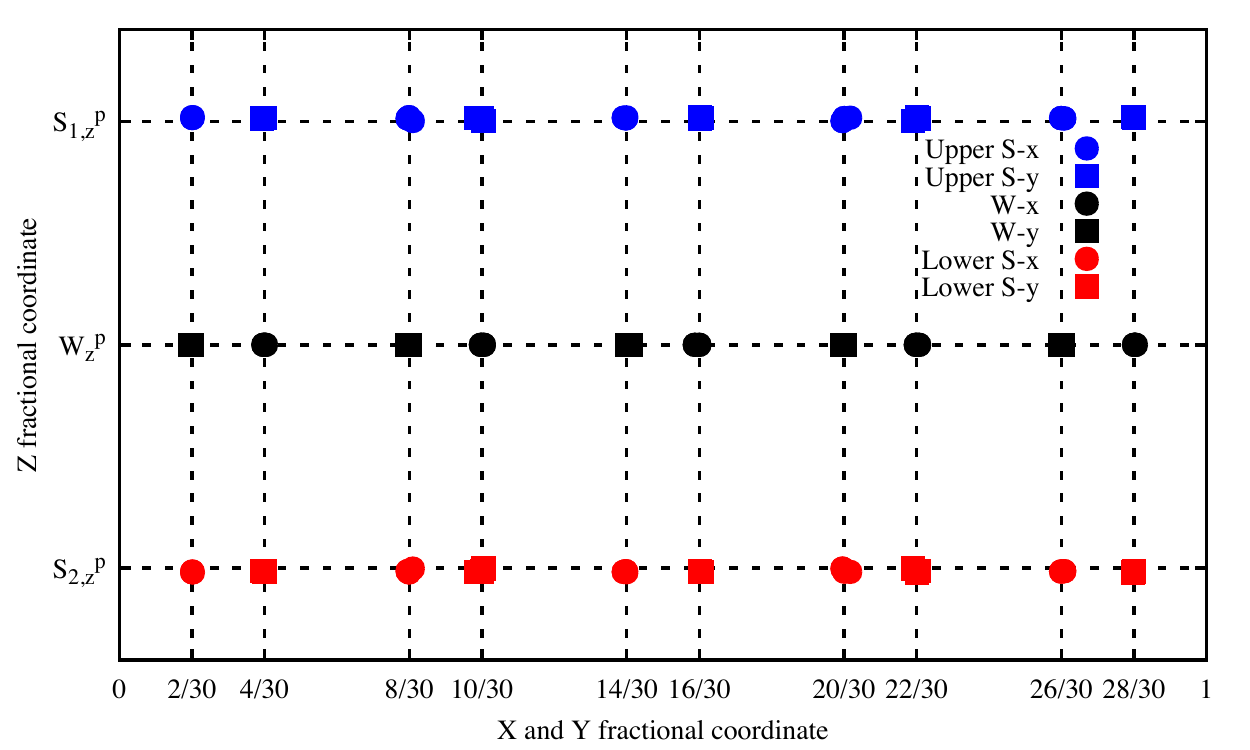}
        \caption{W vacancy.}
        \label{wv55coords}
    \end{subfigure}
        \begin{subfigure}[b]{0.49\textwidth}
        \includegraphics[width=\textwidth]{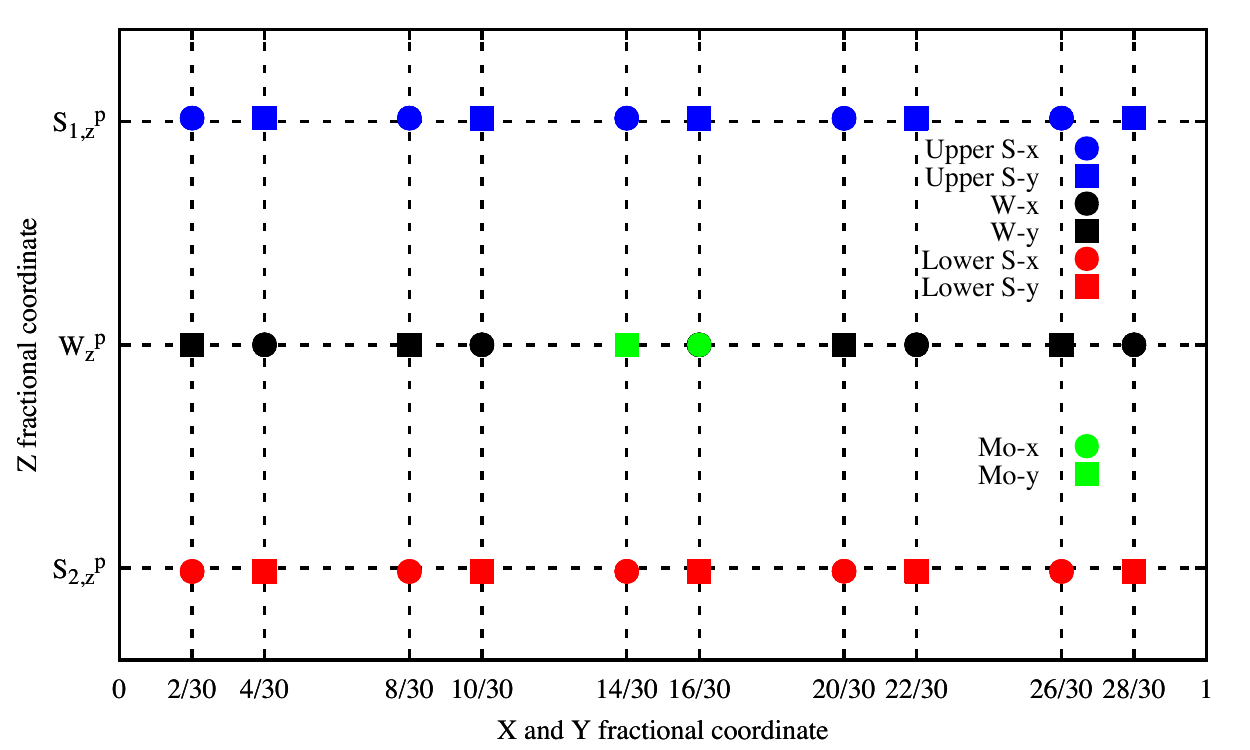}
        \caption{\mow{} substitution.}
        \label{mor55coords}
    \end{subfigure}
        \begin{subfigure}[b]{0.49\textwidth}
        \includegraphics[width=\textwidth]{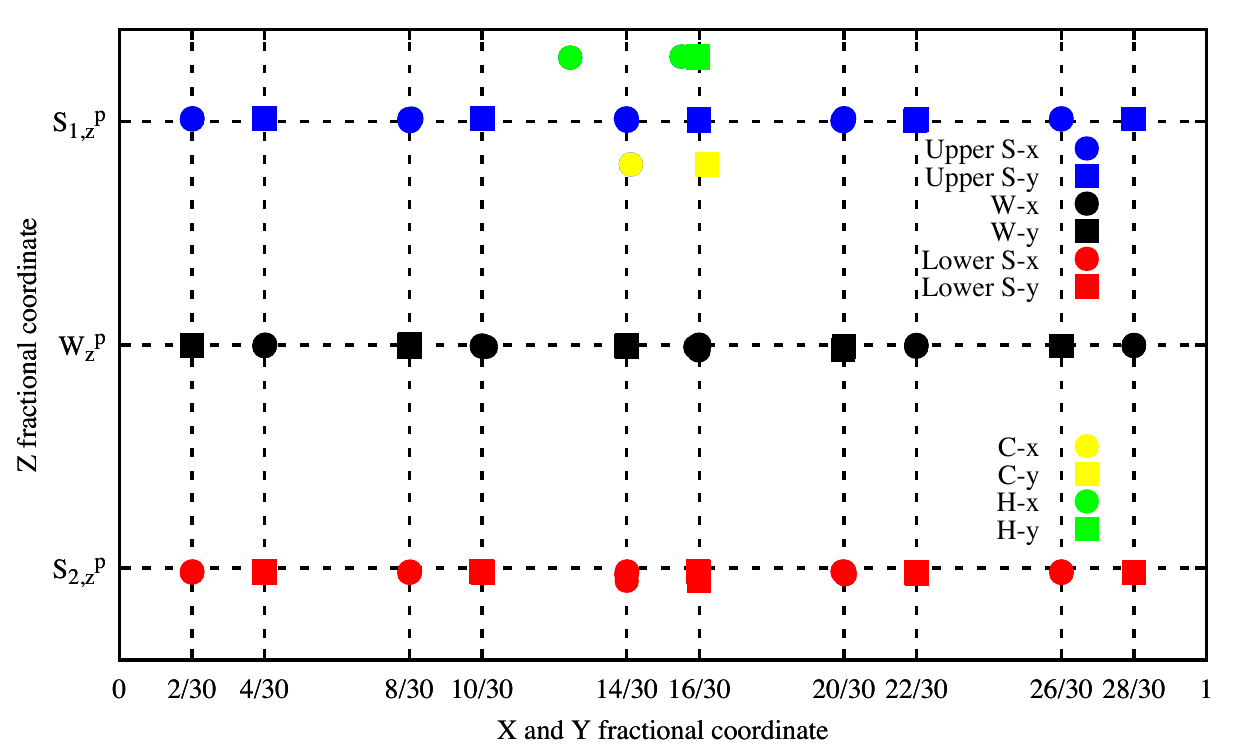}
        \caption{\ch{} substitution.}
        \label{ch255coords}
    \end{subfigure}
    \caption{(color online) Changes in the fractional Z coordinates with the X (circles) and Y (squares) coordinates for the atoms in the \ws{} supercell for each defect.  The grid marks the positions of unrelaxed ions generated with the pristine supercell, with $S_{1,z}^p$, $S_{2,z}^p$, and $W_{z}^p$ being the pristine $z$ coordinates of the top S-layer, the middle W-layer, and the bottom S-layer, respectively. The most significant changes happen in (a) for the S vacancy, where ions near its site move downwards and towards the middle.} 
    \label{fig6}
\end{figure}

We then used the relaxed coordinates from our 3\by3 supercells as a starting point in our 5\by5 runs by surrounding the former in unrelaxed rows and columns of pristine unit cells. In Fig.~\ref{sv55coords} we show the resulting changes in fractional coordinates for the S vacancy. Here an S ion was removed from the top layer in the middle of the supercell. The resulting effect is that the stacking layers are deformed with the ions moving downwards and to the centre as we approach the vacancy site. Structural changes are less perceptible in the other three defects, specially for the W vacancy and the \mow{} substitution, as mirror plane symmetry is preserved. There are, however, some changes in bond lengths. W-S bond lengths change from 2.37 \AA~ near the W-vacancy to 2.420 \AA~ at the supercell's border, meaning that that S-ions move away from the W vacancy's site. For \mow{}, Mo-S bond length is 2.420 \AA, while the W-S bond length changes from 2.415 \AA~ near the substitution to 2.420 \AA~ near the border. Both these defected systems preserve $C_3$ rotational symmetry. 
Finally, for the \ch{} substitution, the H-C-H angle changes from 100.95 \degree to 94.41 \degree , while the C-H bond length changes from 1.12 \AA~to 1.11 \AA~(the first values refer to an isolated CH$_2$ molecule whose geometry was optimised using the same pseudo-potentials). The W-S bond length for the sulfur ion below the methylene molecule varies between 2.415 \AA and 2.425 \AA~and goes back to 2.420 \AA~at the 
edge of the supercell.

To understand which of the four defects is more energetically favourable, we computed their formation energies using
Formation energies are computed using
\be
E_\text{formation} = E_\text{defect}^{5\times5} + E_\text{removed ion} - E_\text{pristine}^{5\times5} - E_\text{added ion} ,
\ee
where $E_\text{defect}^{5\times5}$ and $E_\text{pristine}^{5\times5}$ are the DFT total energies for the 5\by5 defect system and the pristine system, respectively, while $ E_\text{removed ion}$ and $E_\text{added ion}$ are the DFT energies of the ions which are removed or added, respectively (for the vacancies, $E_\text{added ion}=0$). For both \mow{} substitution and W vacancy the reference ion energies used were the bulk systems, while for the \ch{} substitution and the S vacancy we used as reference the isolated molecule for the former and the isolated ion for the latter. The results are shown in Tab.~\ref{tab1}. The most energetically stable defect is the \ch{} substitution (with a formation energy of -0.179 Ry), while the least energetically favourable is the W vacancy which needs 0.797 Ry.

\section{Systems with no mid-gap states: \mow{} and \ch{}}
\begin{figure}[htp]
    \begin{subfigure}[b]{0.49\textwidth}
        \includegraphics[width=\textwidth]{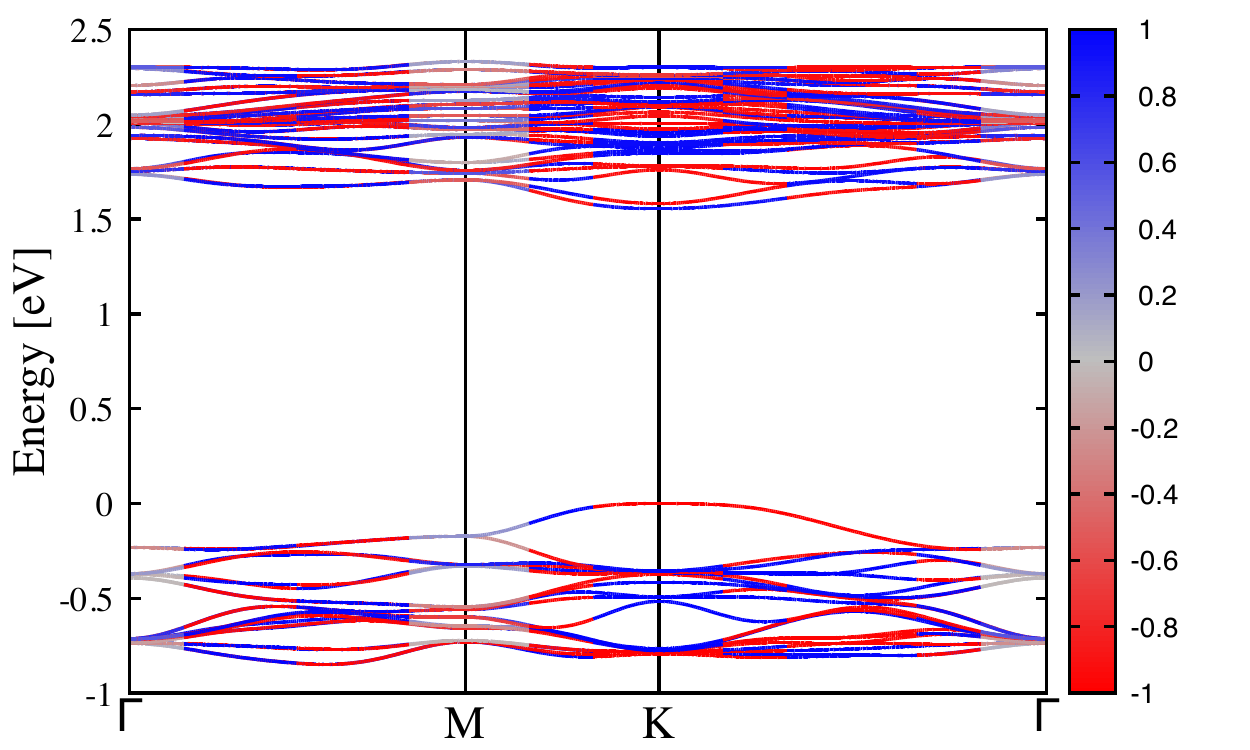}
        \caption{\mow{} substitution.}
        \label{morbands}
    \end{subfigure}
    \begin{subfigure}[b]{0.49\textwidth}
        \includegraphics[width=\textwidth]{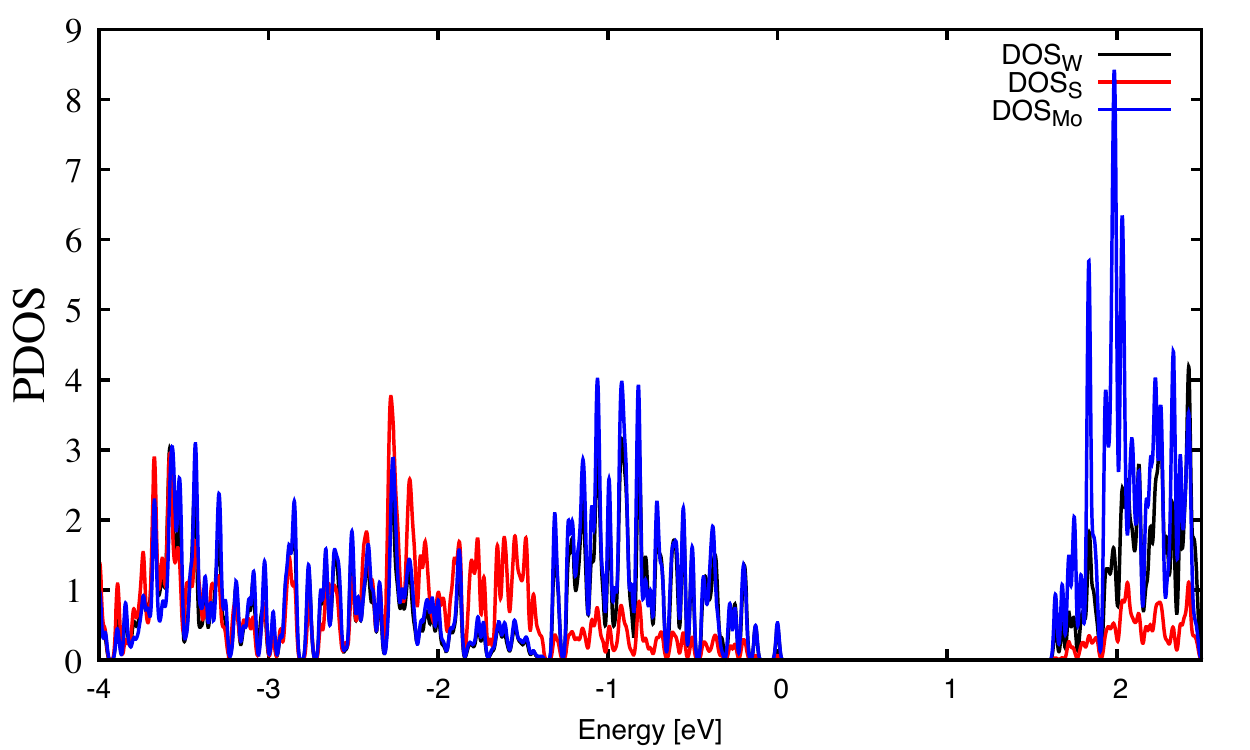}
        \caption{\mow{} substitution.}
        \label{mor55dos}
    \end{subfigure}
    \begin{subfigure}[b]{0.49\textwidth}
        \includegraphics[width=\textwidth]{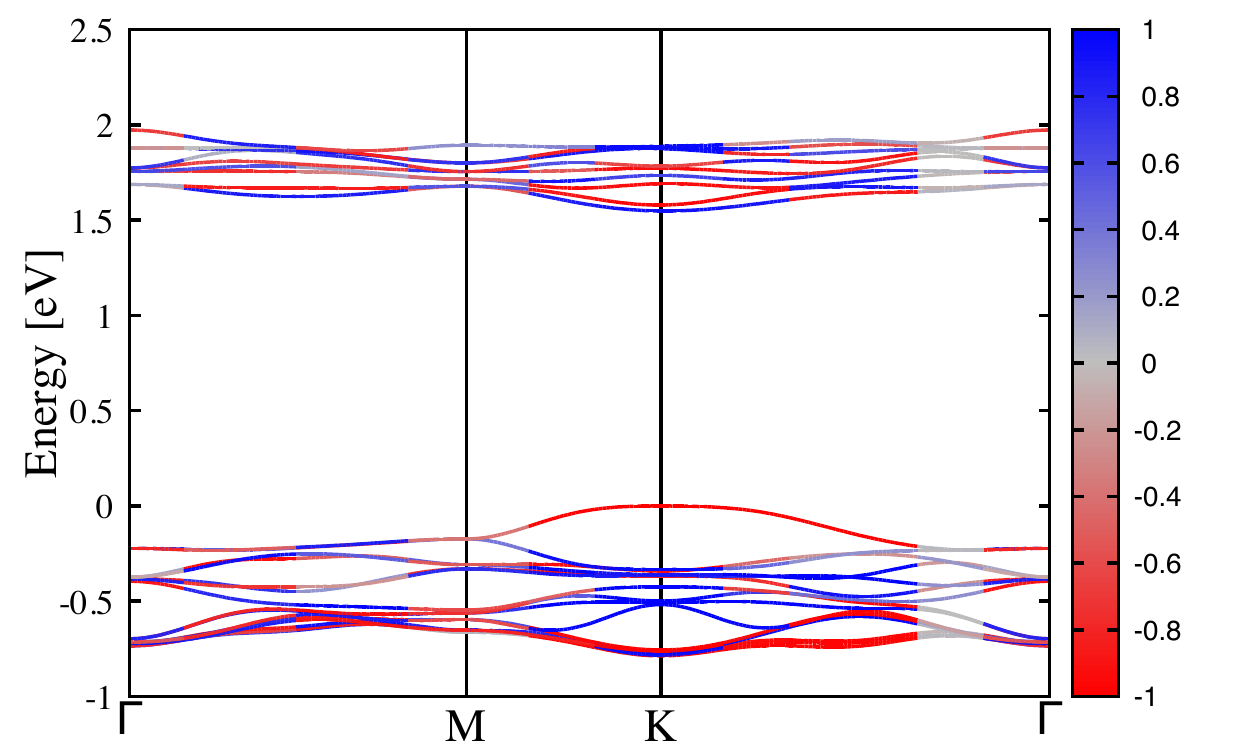}
        \caption{\ch{} substitution.}
        \label{ch2bands}
    \end{subfigure}
    \begin{subfigure}[b]{0.49\textwidth}
        \includegraphics[width=\textwidth]{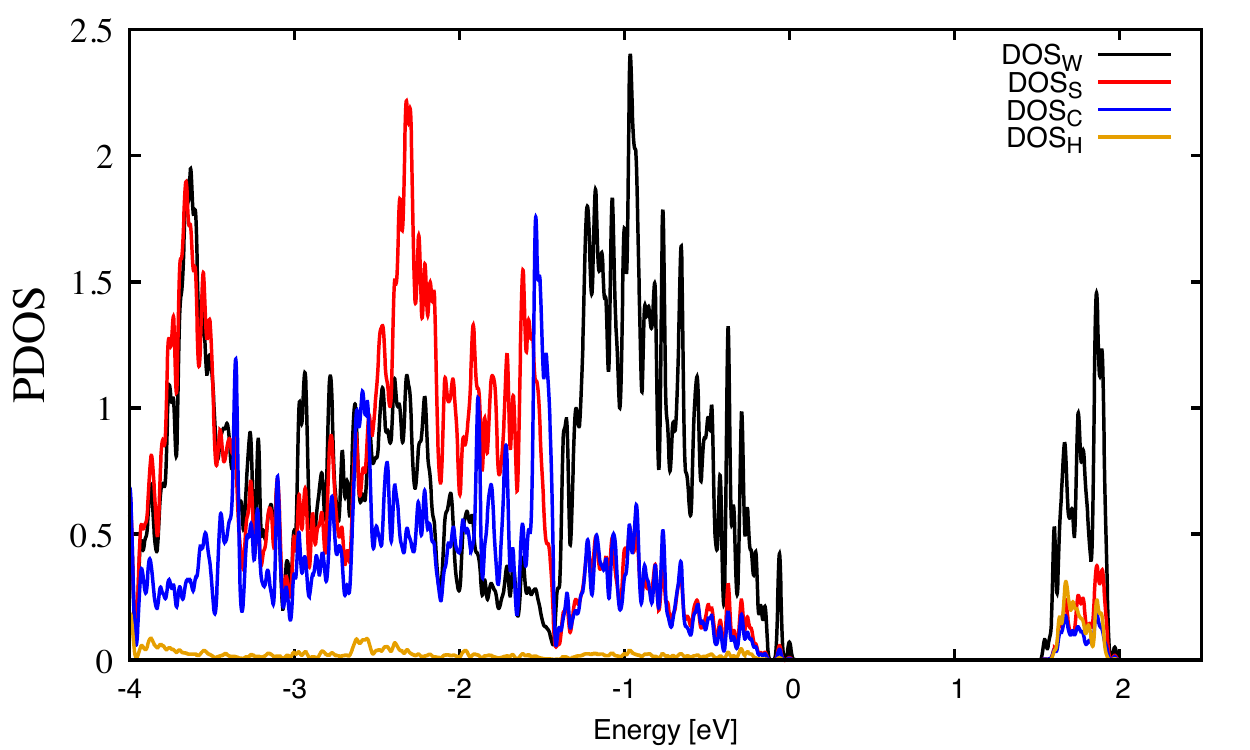}
        \caption{\ch{} substitution.}
        \label{ch255dos}
    \end{subfigure}
    \caption{(color online) DFT band structures and spin-texture of defected \ws{}. a) Mo$_\mathrm{W}$ substitution; b) \ch{} substitution. color scale indicates value of $\braket{S_z}$ (red for -1 and blue for 1).  DFT projected density of states (PDOS) for defected \ws{} systems with no mid-gap states. a) Mo$_\mathrm{W}$ substitution; b) \ch{} substitution. The dominant contributions from to the density of states from the defects occur at energies much lower than those at the transitions that compose the exciton.}
    \label{fig8}
\end{figure}

As shown in the main text, both \mow{} and \ch{} substitutions show no new absorption peaks. At the band structure level they also do not show any mid-gap states, as Figs.~\ref{morbands} and~\ref{ch2bands} show. By plotting the projected density of states in Figs.~\ref{mor55dos} and~\ref{ch255dos} and analysing the peaks corresponding to the Mo ion and \ch molecule, it is possible to see that these show at single particle energies that are much lower that those involved in the formation of the A and B excitons. So if these defects do show up in the spectrum, it must be at emission energies significantly higher than the A and B exciton states. 

\section{Defect bound excitonic wave functions}
\begin{figure}[htpb]
    \begin{subfigure}[b]{0.49\textwidth}
	\includegraphics[width=\textwidth]{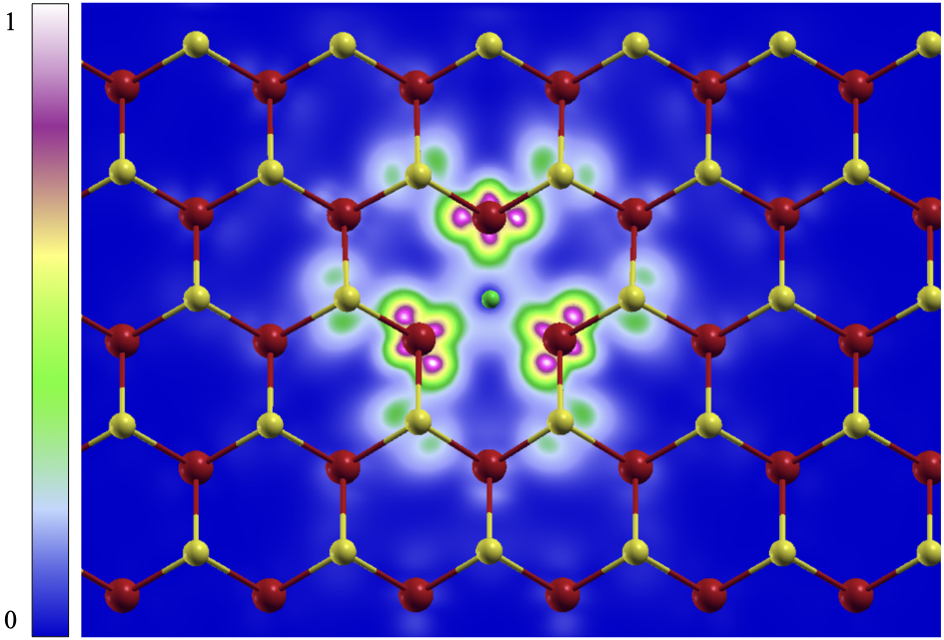}
	\caption{S vacancy \dd{1}'s excitonic wavefunction.}
		\label{svD2}
    \end{subfigure} 
        \begin{subfigure}[b]{0.49\textwidth}
	\includegraphics[width=\textwidth]{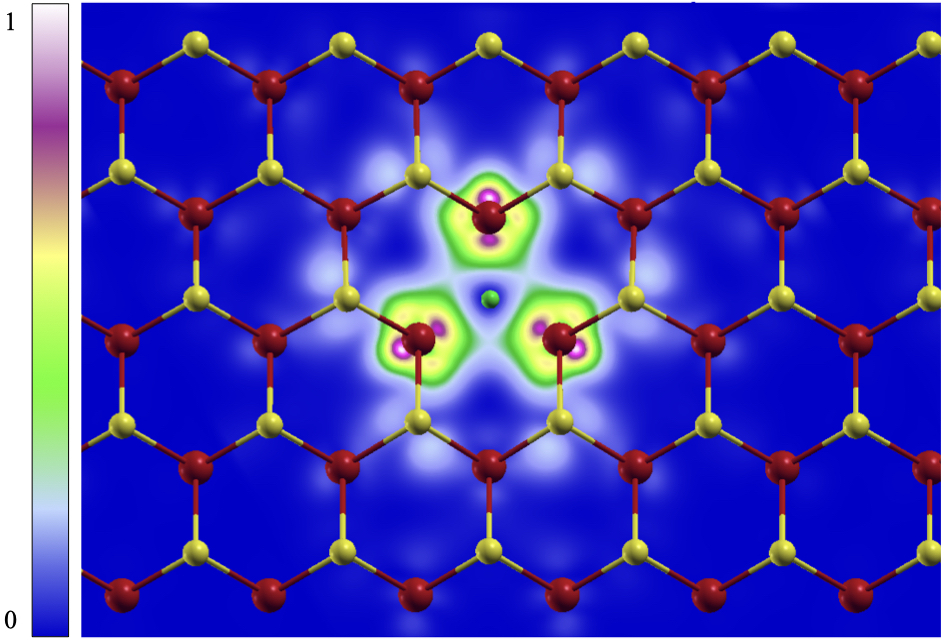}
	\caption{S vacancy \dd{2}'s excitonic wavefunction.}
		\label{svD1}
    \end{subfigure}    
    \caption{(color online) Excitonic wave-functions for all peaks identified in Fig. 3a. The hole's position is marked by the green sphere, at the position of the S vacancy. The colour bar indicates the magnitude of the exctionic wave function's projection on the supercell. For both \dd{1} and \dd{2} both the electron and the hole are localised on the same site.}
    \label{sv55defects-full}
\end{figure}

\begin{figure}[htpb]
    \begin{subfigure}[b]{0.49\textwidth}
	\includegraphics[width=\textwidth]{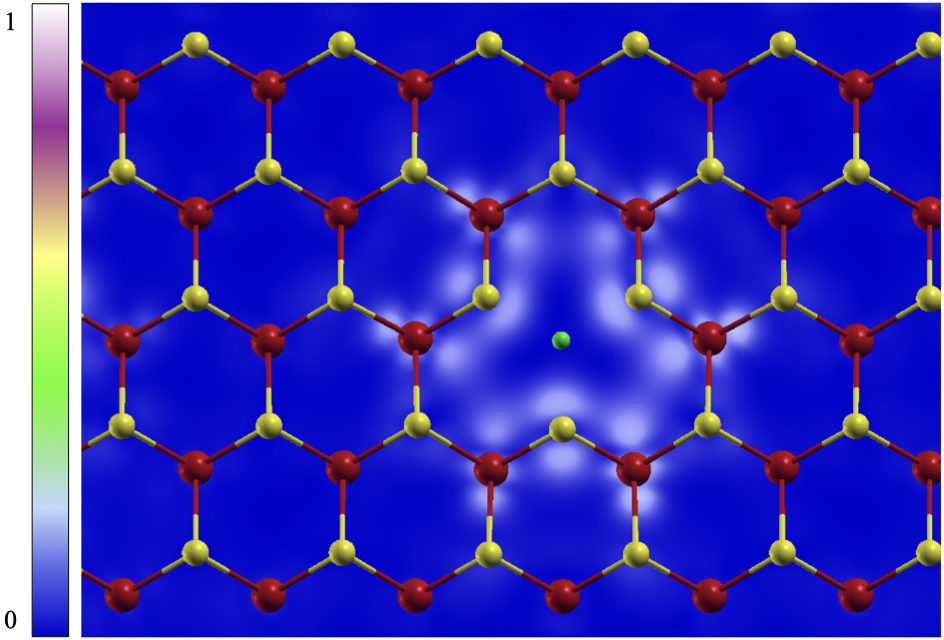}
	\caption{W vacancy \dd{1}'s excitonic wavefunction.}
		\label{wvD1}
    \end{subfigure} 
    \begin{subfigure}[b]{0.49\textwidth}
	\includegraphics[width=\textwidth]{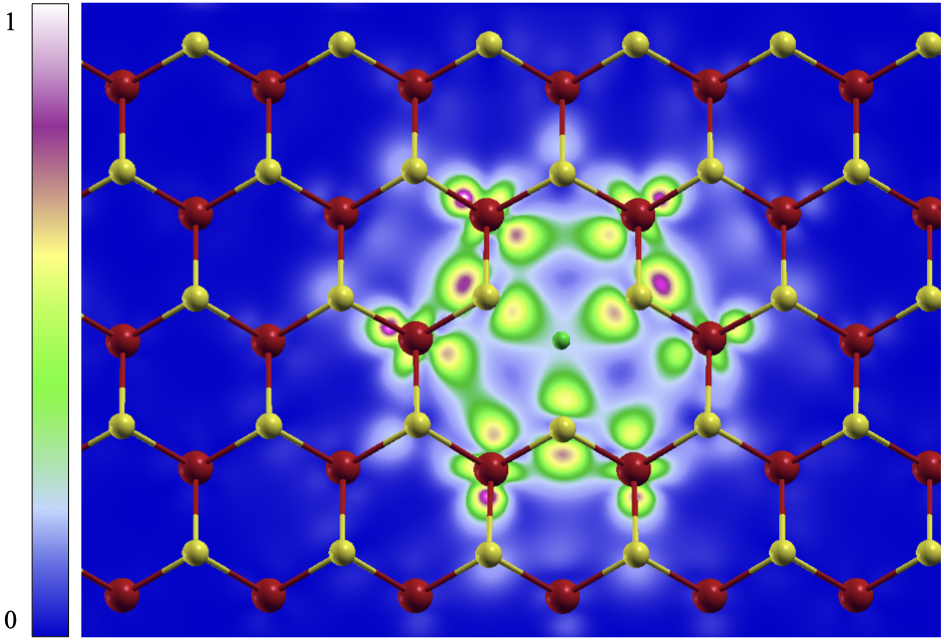}
	\caption{W vacancy \dd{2}'s excitonic wavefunction.}
		\label{wvD2}
    \end{subfigure}    
        \begin{subfigure}[b]{0.49\textwidth}
	\includegraphics[width=\textwidth]{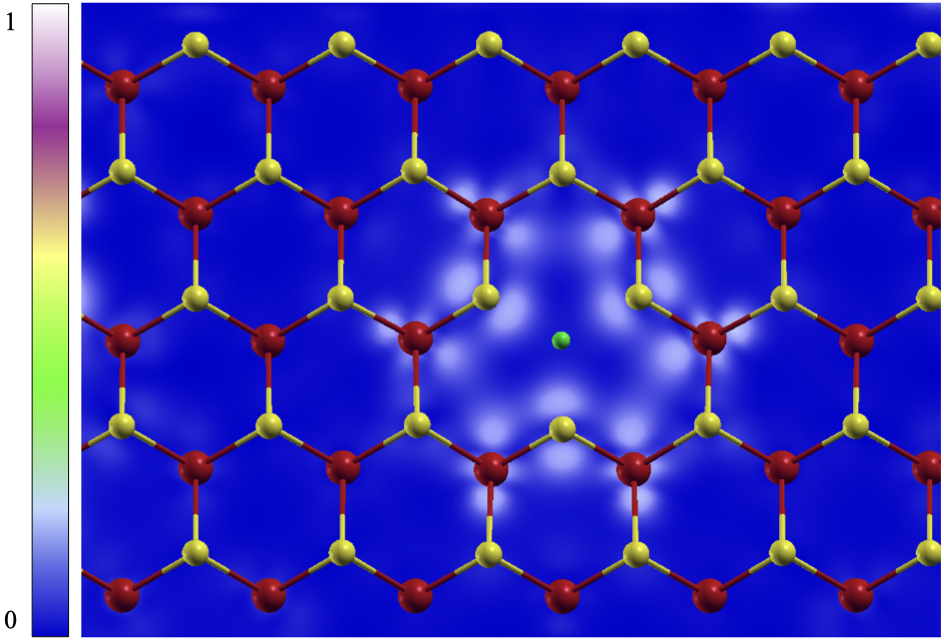}
	\caption{W vacancy \dd{3}'s excitonic wavefunction.}
		\label{wvD3}
    \end{subfigure}    
    \begin{subfigure}[b]{0.5\textwidth}
	\includegraphics[width=\textwidth]{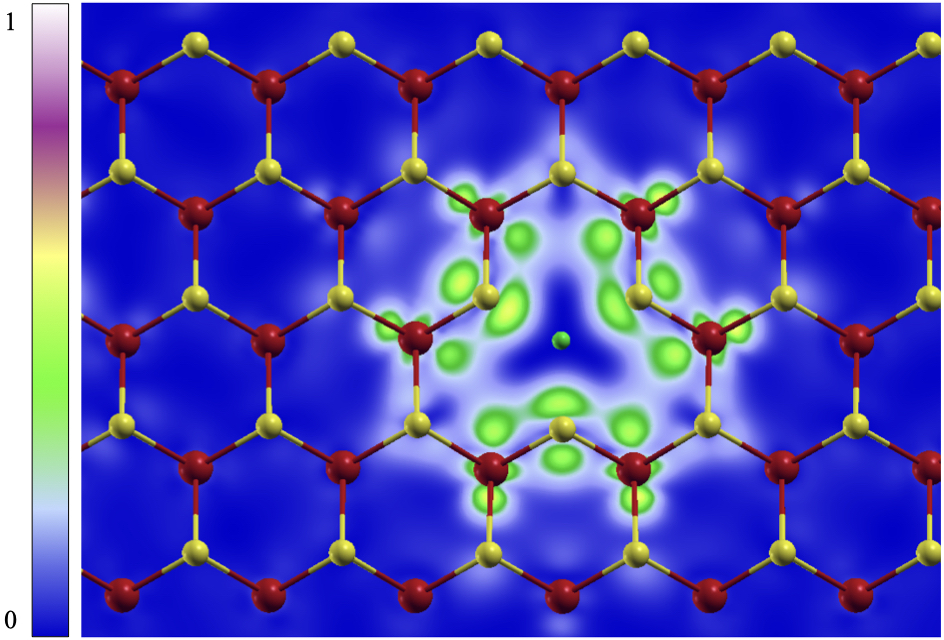}
	\caption{W vacancy \dd{4}'s excitonic wavefunction.}
		\label{wvD4}
    \end{subfigure}    
    \begin{subfigure}[b]{0.5\textwidth}
	\includegraphics[width=\textwidth]{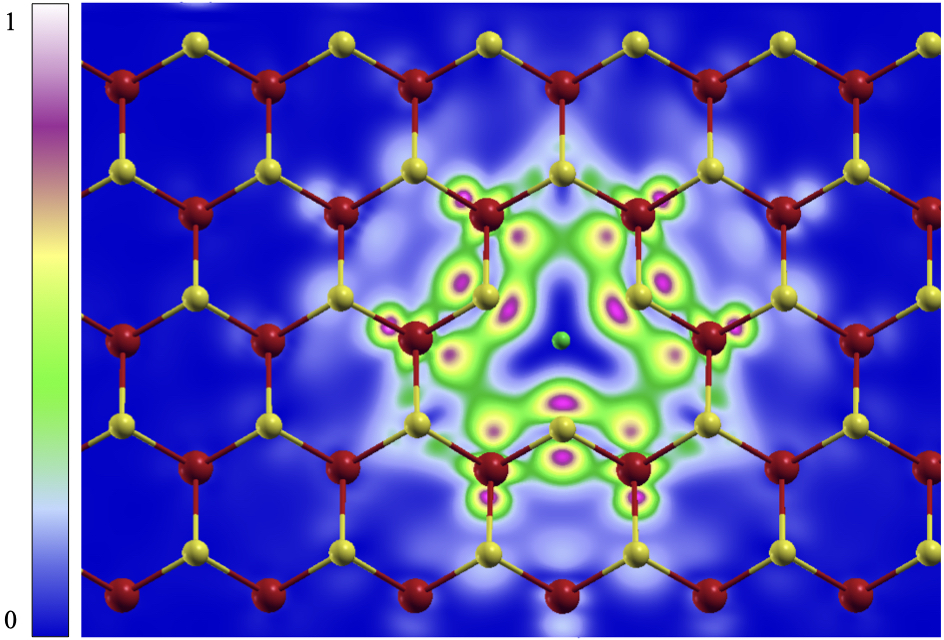}
	\caption{W vacancy \dd{5}'s excitonic wavefunction.}
		\label{wvD5}
    \end{subfigure} 
    \caption{(color online) Excitonic wave-functions for all peaks identified in Fig. 3b. The hole's position is marked by the green sphere, at the position of the W vacancy. The colour bar indicates the magnitude of the exctionic wave function's projection on the supercell. In (a) and (c) the electron part of the wave function is almost entirely not localised at the vacancy site, resulting from an interaction with charges in adjacent repetitions of the supercell. For (b), (d), and (e) both electron and hole are localised on the same site.}
    \label{wv55defects-full}
\end{figure}

In Figs.~\ref{sv55defects-full} to~\ref{wv55defects-full} we show the projection of each exciton wave function on the supercell, with the hole placed at the defect site. The two defect bound excitonic states that we identified in our calculations (Figs.~\ref{svD1} and~\ref{svD2}) are completely localised near the vacancy, where the hole is, indicating that these are truly new defect-bound absorption peaks. 

\begin{figure}[htpb]
    \begin{subfigure}[b]{0.5\textwidth}
	\includegraphics[width=\textwidth]{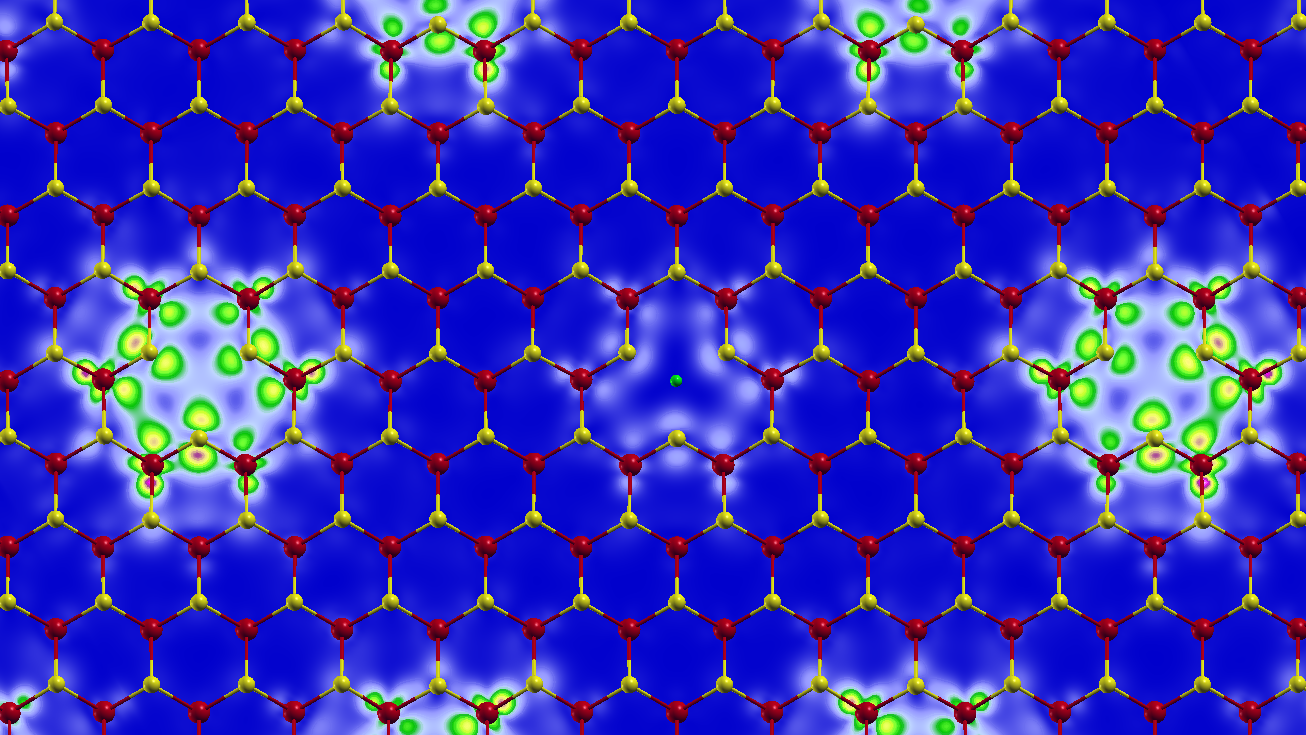}
	\caption{Zoom out of W vacancy \dd{1}'s excitonic wavefunction.}
		\label{wvD1-zoom}
    \end{subfigure} 
    \begin{subfigure}[b]{0.5\textwidth}
	\includegraphics[width=\textwidth]{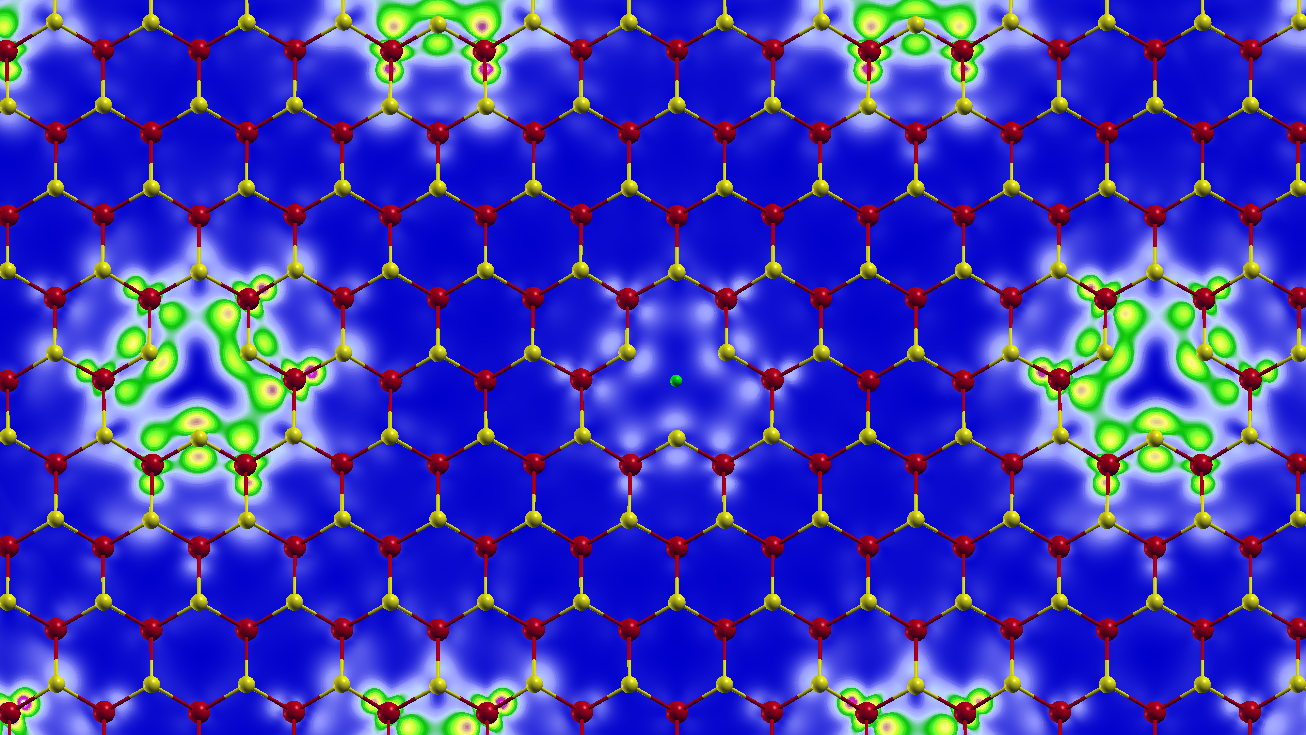}
	\caption{Zoom out of W vacancy \dd{3}'s excitonic wavefunction.}
		\label{wvD3-zoom}
    \end{subfigure}    
    \caption{(color online) Excitonic wave-functions for all peaks identified in Fig. 3b. The hole's position is marked by the green sphere, at the position of the W vacancy. While there is some probability density near the vacancy site, most of the wave function is localised in nearby vacancies.}
    \label{wv55-zoom}
\end{figure}

The projections of the defect associated exciton wave functions for the W vacancy are shown in Fig.~\ref{wv55defects-full}, while in Fig.~\ref{wv55-zoom} we show a zoomed out projection of exciton states \dd{1} and \dd{3}. This is to emphasise the fact that both \dd{1} and \dd{3} result from the spurious interaction between a hole and an electron at different vacancy sites. At larger supercells these states should vanish from the absorption spectrum. 

\begin{table}[htp]
    \centering
    \caption{Exciton energies for each system (the exciton ID are the same as the ones shown in Figs. 3 and 4), optical strength ratio between the A exciton of a given system and the remaining others, and respective formation energies.}
    \begin{tabular}{ccccccccc}
        \toprule
        System   &   \multicolumn{7}{c}{Energy [eV] (intensity ratio to A exciton)}  & Formation energy [eV]\\ 
                       &  A               & B    & \dd{1} & \dd{2} & \dd{3} & \dd{4} & \dd{5} & \\
        \midrule
        Pristine      & 1.83 & 2.21 &     -      &  -        &   -        &   -       &   -        &  - \\
        S vacancy & 1.83  & 2.16  & 1.35   & 1.14   &    -      &    -       &     -      & 2.76196\\
        W vacancy & 1.86  & 2.17 & 1.18  & 0.98  & 0.93 & 0.67  & 0.58   & 10.8437\\
        \mow{}    & 1.89  & 2.24 &    -    &    -    &    -    &    -    &   -     & 0.312931 \\
        \ch{}        & 1.84 & 2.18 &    -    &   -     &     -   &  -      &    -    & -2.43542\\
        \bottomrule
    \end{tabular}
    \label{tab1}
\end{table}


\section{Methodology}
\subsection{DFT}
Geometry optimisation and ground state calculations were performed with the Quantum Espresso package. Supercells where generated by replicating the pristine cell (previously opimised) and then introducing the respective defect. The out of plane direction was isolated from periodic boundary conditions using a Coulomb cutoff technique from Rozzi et al.\cite{Rozzi2006}. A plane wave kinetic energy cutoff of 95 Ry was used, with a tolerance parameter of 10$^{-12}$ for the error in the density. For geometry optimisation, we converged the forces down to 10$^{-5}$ Ry/bohr and set the minimum difference in total energies to 10$^{-8}$ Ry. Since properties of the pristine systems where computed on a 30$\times$30 k-point mesh, in the \fbf supercells we used a mesh of 6$\times$6 k-points.

\subsection{\gw}
To correct the energy levels coming from the GGA DFT calculation, \gw calculations were performed, using the Plasmon-pole approximation with the \yambo code. A total of 1700 bands where necessary to converge the screening, while kinetic cut-offs of 2 Ha where used in the integrals. This allowed for convergence within the range of the milli-electron-volt. The final result was a scissor operator which was used in the computation of the optical spectrum.

\subsection{BSE}
We computed the absorption spectra with the Bethe-Salpeter equation (BSE)~\cite{Onida2002} as implemented in the \yambo code~\cite{Sangalli2019}, to properly account for excitonic effects. This involves solving the eigenvalue equation

\begin{equation}
    (\varepsilon_\ck^\text{GW}  -\varepsilon_\vk^\text{GW}) A_{vc\kk}^\gl + \sum_{\kk' c' v'}K^{eh}_{\substack{vc\kk \\ v'c'\kk'}}A_{v'c'\kk'}^\gl = E_\gl A_{vc\kk}^\gl,
    \label{eq1}
\end{equation}
where $\varepsilon_{\ck/\vk}$ are quasi-particle band energies obtained from the \gw COHSEX approximation~\cite{Onida2002} and $K^{eh}$ accounts for the static screened and bare Coulomb interactions between electrons and holes. The excitonic wave function, $\ket{\gl}$, is then constructed by using
\be
\ket{\gl} = \sum_{\kk,c,v}A_{vc\kk}^\gl\ket{c\kk}\ket{v\kk}.
\ee

In our calculations we focus on the bound exciton energies below the band gap, so the BSE kernel only has the electron-hole pairs needed to converge the pristine system's first two excitons, marked as A and B. The optical absorption is then obtained directly from the imaginary part of the dielectric function, $\varepsilon(\go)$, which is given (within the Tamm-Dancoff approximation) by

\begin{equation}
    \varepsilon(\go) = -2 \lim_{\qq \to 0}\sum_{\gl,c,v,\kk} \frac{\left|\braket{v\kk-\qq|e^{-i\qq\cdot\rr}|c\kk}A_{cv\kk}^\gl\right|^2}{\go - E_\gl + i\eta}.
    \label{eq2}
\end{equation}
Here $c$/$v$ represent conduction/valence states, $\kk$ is the k-point vector in the Brillouin zone. $\gl$, $E_\gl$, and $A_{cv\kk}^\gl$ are the exciton index, eigenenergy and eigenvector of Eq.~\ref{eq1}. The electron-hole interaction is included in $K^{eh}$, which contains both the un-screened exchange interaction and the screened direct Coulomb interaction. For the screening and cut-offs, we took the same values as the ones that were used in the \gw runs. In order to properly account for the bands in the same energy range as the pristine case, we had to use a total of 60 bands in the BSE kernel.

\bibliographystyle{MSP}
\bibliography{literature}